\documentclass[useAMS,usenatbib,usegraphicx]{mn2e}
\usepackage{times}

\usepackage{graphicx}
\usepackage{epsfig}
\usepackage{amssymb}
\usepackage{natbib}
\usepackage{journals}
\usepackage[latin1]{inputenc}
\newcommand{\Sauron}{\texttt{SAURON}}
\newcommand{\XSauron}{\texttt{XSauron}}
\newcommand{\kms} {$\mbox{km s}^{-1}$}

\def\spose#1{\hbox to 0pt{#1\hss}}
\def\lta{\mathrel{\spose{\lower 3pt\hbox{$\sim$}}
    \raise 2.0pt\hbox{$<$}}}
\def\gta{\mathrel{\spose{\lower 3pt\hbox{$\sim$}}
    \raise 2.0pt\hbox{$>$}}}
\newdimen\hssize
\newdimen\hdsize
\def\VV{V_\mathrm{RMS}}
\title[The \Sauron\ project -- IX]
 {The SAURON project -- IX. A kinematic classification for early-type galaxies.}
\author[E.\ Emsellem et al.]
{Eric Emsellem,$^1$ 
  Michele Cappellari,$^{2,3}$
  Davor Krajnovi{\'c},$^3$ 
  Glenn van de Ven,$^{2,4,5}$\thanks{Hubble Fellow}
  R.\ Bacon,$^1$ 
\newauthor  M.\ Bureau,$^3$ 
  Roger L.\ Davies,$^3$ 
  P.\ T.\ de Zeeuw,$^2$
  Jes\'us Falc\'on-Barroso,$^{2,6}$
\newauthor  Harald Kuntschner,$^7$ 
  Richard McDermid,$^2$ 
  Reynier F.\ Peletier,$^{8}$
  Marc Sarzi,$^9$ \\
$^1$Université de Lyon, France; université Lyon 1, F-69007; CRAL, Observatoire
de Lyon, F-69230 Saint Genis Laval; CNRS, UMR 5574 ; ENS de Lyon, France \\
$^2$Sterrewacht Leiden, Leiden University, Niels Bohrweg~2, 2333~CA Leiden, The Netherlands\\
$^3$Sub-Department of Astrophysics, University of Oxford, Denys Wilkinson Building, Keble Road, Oxford OX1 3RH, United Kingdom \\
$^4$Department of Astrophysical Sciences, Peyton Hall, Princeton, NJ 08544, USA \\ 
$^5$Institute for Advanced Study, Einstein Drive, Princeton, NJ 08540, USA \\
$^6$European Space and Technology Centre (ESTEC), Keplerlaan 1, Postbus 299, 2200 AG Noordwijk, The Netherlands \\
$^7$Space Telescope European Coordinating Facility, European Southern Observatory, Karl-Schwarzschild-Str~2, 85748 Garching, Germany\\
$^8$Kapteyn Astronomical Institute, Postbus 800, 9700 AV Groningen, The Netherlands \\
$^9$Centre for Astrophysics Research, University of Hertfordshire, Hatfield, Herts AL10 9AB }
\begin{document}
\maketitle
%
%
\begin{abstract}
Two-dimensional stellar kinematics of 48 representative E and S0 galaxies obtained with the
\Sauron\ integral-field spectrograph reveal that
early-type galaxies appear in two broad flavours, depending on
whether they exhibit clear large-scale rotation or not. 
We define a new parameter $\lambda_R \equiv \langle R \, \left| V \right| \rangle / 
\langle R \, \sqrt{V^2 + \sigma^2}\rangle $, which involves luminosity weighted averages over
the full two-dimensional kinematic field, as a proxy to quantify
the observed projected stellar angular momentum per unit mass. 
We use it as a basis for a new kinematic 
classification: early-type galaxies are separated into slow and fast rotators,
depending on whether they have $\lambda_R$ values within 
their effective radius $R_e$ below or above 0.1, respectively.
Slow and fast rotators are shown to be physically distinct classes of galaxies, a result which
cannot simply be the consequence of a biased viewing angle. 
Fast rotators tend to be relatively low luminosity galaxies with $M_B \ga -20.5$.
Slow rotators tend to be brighter and more massive galaxies, but are still spread over a 
wide range of absolute magnitude. Three slow rotators of our sample, among the most massive ones, 
are consistent with zero rotation. Remarkably, all other slow rotators 
(besides the atypical case of NGC~4550) contain a large kpc-scale kinematically decoupled core (KDC). 
All fast rotators (except one galaxy with well-known irregular shells)
show well aligned photometric and kinemetric axes, 
and small velocity twists, in contrast with most slow rotators which exhibit significant 
misalignments and velocity twists.
These results are supported by a supplement of 18 additional early-type
galaxies observed with \Sauron. In a companion paper (Paper~X), 
we also show that fast and slow rotators are distinct classes
in terms of their orbital distribution. We suggest that gas is a key ingredient 
in the formation and evolution of fast rotators, and that the slowest rotators are
the extreme evolutionary end point reached deep in gravitational potential wells where dissipationless
mergers had a major role in the evolution, and for which most of the baryonic angular
momentum was expelled outwards. Detailed numerical simulations in
a cosmological context are required to understand how to form large-scale kinematically
decoupled cores within slow rotators, and more generally to explain the distribution
of $\lambda_R$ values within early-type galaxies and the 
distinction between fast and slow rotators.
      
\end{abstract}
\begin{keywords}
galaxies: elliptical and lenticular, cD~--
galaxies: evolution~-- 
galaxies: formation~-- 
galaxies: kinematics and dynamics~-- 
galaxies: structure
\end{keywords}
%
%
\section{Introduction\label{sec:intro}}

The origin of the classification fork for galaxies
can be found in an early paper by Jeans (1929), with the S0s as a class being
introduced by \cite{Hubble36} to account for the important population
of flattened objects in nearby clusters \citep{SB51}. In a recent debate on
galaxy classification, \cite{Sandage04} mentioned that the simplest definition of an S0
galaxy remains ``a disc galaxy more flattened than an E6 elliptical but with no
trace of spiral arms or recent star formation".
Elliptical (E) and lenticular (S0) galaxies are usually 
gathered into the so-called early-type category, and are recognised to share 
a number of global properties \citep{RC3} such as their relatively low dust 
and interstellar gas content and their overall red colours. 
The Hubble sequence is however seen as a continuous one between ellipticals
and spirals, with the S0s occupying the transition region with typical bulge to
disc ratios of $\sim 0.6$. S0s are thus considered disc-dominated galaxies, 
while Es are spheroid-dominated. 

Such contrived galaxy types may be misleading, most evidently because ``the sequence E0-E6
is one of apparent flattening" \citep[][hereafter KB96]{KB96}. A modern classification scheme
should go beyond a purely descriptive tool, and should therefore
encompass part of our knowledge of the physical properties of these 
objects\footnote{For the written transcript of a debate on this issue, see 
\cite{Sandage04} and the panel discussion in the
same Proceedings.}. This was advocated by KB96 
who wished to update the Hubble sequence by sorting ellipticals in terms of 
the importance of rotation for their stellar dynamical state.
They used the disciness or boxiness of the isophotes
to quantify anisotropy and to define refined types: E(d) galaxies (for discy ellipticals) making the link
between S0s and E(b) galaxies (for boxy ellipticals). The
disciness (or boxiness) was then provided by a measure of the now classical
normalised $a_4/a$ term \citep[see][for details]{BDM88}: 
positive and negative $a_4 / a$ terms correspond to discy and boxy deviations
from ellipses, respectively. 

This extension of the Hubble types has the merit of upgrading our view
of Es and S0s via some easily accessible observable parameter, and
it follows the philosophy that a mature classification scheme should include 
some physics into the sorting criteria. 
It does, however, use a photometric indicator as an attempt to quantify
the dynamical state of the galaxy, which may be unreliable.
More importantly, it conserves the dichotomy between S0s and Es, relying
on the old (and ambiguous) definition of an S0.

We have recently conducted a survey of 72 early-type (E, S0, Sa) galaxies using the
integral-field spectrograph \Sauron\ mounted on the William Herschel Telescope
in La Palma (\citealt{Bacon+01}, hereafter Paper~I; \citealt{PaperII}, hereafter Paper~II). 
This allowed us to map the stellar and gas
kinematics as well as a number of stellar absorption line indices up
to about one effective radius $R_e$ for most of the galaxies in the sample.
The two-dimensional stellar kinematics for the 48 E and S0 galaxies 
\citep[][hereafter Paper~III]{Emsellem+04} show a wide 
variety of features such as kinematically decoupled
or counter-rotating cores, central discs and velocity twists.
More importantly, there seem to be two broad classes of observed stellar
velocity fields, with galaxies in one class exhibiting a clear large-scale rotation
pattern and those in the other showing no significant rotation
(Fig.~\ref{fig:allV}). The existence of these two classes must be linked
to the formation and evolution of early-type galaxies, and is in any case 
a key to understand their dynamical state \cite[see also][hereafter Paper~IV]{PaperIV}.

Using the unique data set obtained in the course of the \Sauron\ 
project, we here revisit the early-type galaxy classification 
issues mentioned above, using the available full two-dimensional kinematic information.
A companion paper \cite[][hereafter Paper~X]{PaperX} examines in more detail the 
orbital anisotropy of elliptical and lenticular galaxies using the same data set.
After a brief presentation of the data set and methods (Sect.~\ref{sec:data}), 
we define a new parameter as a proxy to robustly quantify 
the angular momentum of galaxies (Sect.~\ref{sec:LambdaR}). 
In Section~\ref{sec:Hubble}, we examine how this parameter relates both 
to more standard photometric classification schemes including
de Vaucouleurs morphological classification of Es and S0s \citep{RC3}, 
or the revision of the Hubble sequence by KB96, and kinemetric properties
of early-type galaxies (Sect.~\ref{sec:kinresults}).
We then briefly discuss the implications of our results on the 
potential scenarios for the formation and evolution of these galaxies (Sect.~\ref{sec:discussion}), 
and conclude in Section~\ref{sec:conclusions}.

%
%

\section{Data and methods} 
\label{sec:data}

\subsection{The \Sauron\ sample of E and S0s}
\label{sec:sample}
The \Sauron\ sample has been designed as a {\em representative}
sample of 72 nearby ($cz < 3000$\kms) early-type galaxies in the plane
of ellipticity $\epsilon$ versus absolute $B$ band magnitude $M_B$.
We restricted the sample to 24 objects for each of the E, S0 and Sa
classes, with 12 `cluster' and 12 `field' targets in each group. 
Although this sample is itself not a complete one, it has been
drawn in a homogeneous way from a complete sample of more than 300 galaxies (see
Paper~II for more details about the full \Sauron\ sample).
We will therefore solely focus on the \Sauron\ 
E$+$S0 sample (which thus contains 48 galaxies) for conducting our analysis in 
Sections~\ref{sec:data} to~\ref{sec:Hubble}. 
An additional set of 18 E$+$S0 galaxies has been independently observed with 
a similar \Sauron\ setup in the course of various other projects. These objects
will be treated as ``specials", most having features which motivated 
a specific observation, and will be only briefly mentioned in Sect.~\ref{sec:discussion}
to confront the obtained results.

\subsection{Photometry}
\label{sec:phot}
Ground-based photometric MDM Observatory data (Falc{\'o}n-Barroso et al. in preparation) 
were obtained for all galaxies of the \Sauron\ sample. We also made use of 
additional Hubble Space Telescope WFPC2 data which are available for 42 galaxies out of the 48 E/S0. 
\Sauron\ reconstructed images were used as well to directly derive
a global ellipticity $\epsilon$ via moments of the 
surface brightness images as:
\begin{equation}
\epsilon \equiv 1 -\sqrt{  \frac{\langle y^2 \rangle}{\langle x^2 \rangle}  } \, ,
\end{equation}
where $x$ and $y$ are the sky coordinates along the photometric 
major and minor axes respectively (see Paper~X for details). 
The brackets $\langle \, \rangle$ correspond to a flux weighted sky average as in:
\begin{equation}
\label{eq:brackets}
\langle G \rangle = \frac{\sum_{i=1}^{N_p} F_i \, G_i}{\sum_{i=1}^{N_p} F_i} \, ,
\end{equation}
where the sums extends over a predefined region on the sky, 
with $N_p$ the number of pixels, and $F_i$ the flux value within the $i^{\mbox{\tiny th}}$ pixel.
In the following Sections, we denote $\epsilon$ and $\epsilon_{1/2}$ 
as the global ellipticity computed from the \Sauron\ data within 1~$R_e$ and $R_e/2$ or
restricted to the equivalent effective aperture of the \Sauron\ field of view,
whichever is smaller. These will be used in Sect.~\ref{sec:LambdaR}, where we examine kinematic quantities 
derived from \Sauron\ data as a function of the flux weighted global ellipticity.

We also obtained radial profiles for the ellipticity and $a_4 / a$ 
(with $a$ the ellipse semi major-axis) parameters 
from the MDM and HST/WFPC2 photometric data sets using the 
{\sc GALPHOT} package \citep{Franx+89}. We then derived $\epsilon_e$
and $(a_4 / a)_e$, the mean ellipticity and $a_4 / a$  
values within 1~$R_e$, by taking the flux weighted average of the {\sc GALPHOT} profiles taking into
account the corresponding sky area. More specifically, the mean $G_e$ of a quantity $G(R)$
within 1~$R_e$ derived from its radial profile, where $R$ is the semi major-axis radius, 
is defined as \citep[see e.g.][]{Ryden+99} 
\begin{equation}
G_e = \frac{\int_0^{R_e} q(R) F(R)\,  G(R) \, R \, \mbox{d}R}{\int_0^{R_e} q(R) F(R)\, R \,  \mbox{d}R}
 \, ,
\end{equation}
where $q(R)$ and $\Sigma(R)$ are respectively the best-fit ellipse axis ratio and 
surface brightness profiles.
Using the sampled radial profiles, we approximate this with
\begin{equation}
G_e \sim \frac{\sum_{k=1}^N q(R_k) F(R_k) G(R_k)\, (R_{\mbox{\small out},k}^2 - R_{\mbox{\small in},k}^2) }
{\sum_{k=1}^N q(R_k) F(R_k)\, (R_{\mbox{\small out},k}^2 - R_{\mbox{\small in},k}^2) } \, ,
\end{equation}
where $R_{\mbox{\small in},k}$ and $R_{\mbox{\small out},k}$ correspond to the inner and 
outer radii of the $k^{\mbox{\tiny th}}$ annulus.
The values for $\epsilon$, $\epsilon_e$ and $(a_4 / a)_e$
are provided in Table~\ref{tab:param}. 
Observed differences between $\epsilon$ and $\epsilon_e$ values are almost
always smaller than 0.1, and due to their respective luminosity weighting.
$\epsilon$ is measured directly from the \Sauron\ data, and will therefore 
be confronted with other quantities measured from the same data set, while
$\epsilon_e$ being derived from one-dimensional radial profiles
can serve in comparison with previous works. 
For both $\epsilon_e$ and $(a_4 / a)_e$, the overall agreement between our values 
and published ones \citep{BSG94} is excellent (see Paper~X for a detailed comparison).

Results from fitting Sersic and Sersic-core laws \citep{Trujillo04+, ACSVI} 
to the radial luminosity profiles will be presented in detail
in a subsequent paper of this series (Falc{\'o}n-Barroso et al., in preparation).
In the present paper, we will only mention trends (Sect.~\ref{sec:sersic}), considering
the Sersic index $n$ (where $n = 1$ corresponds to an exponential luminosity profile, 
and $n=4$ to a de Vaucouleurs $R^{1/4}$ law), as well as the classification of
the central photometric profiles with either shallow or steep inner cusps,
labelled respectively as ``cores" and ``power-laws" \citep{Faber+97,Rest+01,Ravindranath+01,Lauer+05}. 

\subsection{The \Sauron\ data}

\Sauron\ is an integral-field spectrograph built at Lyon Observatory
and mounted since February 1999 at the Cassegrain focus of the William
Herschel Telescope. It is based on the \texttt{TIGER} concept \citep{Bacon+95},
using a microlens array to sample the field of view. Details of the instrument
can be found in Paper~I and II. All 48 E and S0 galaxies were observed with 
the low resolution mode of \Sauron\ which covers a field of view 
of about 33\arcsec$\times$41\arcsec with 0\farcs94$\times$0\farcs94 per square lens. 
Mosaicing was used to cover up to a radius of 1 $R_e$. Only for the two galaxies with the largest 
$R_e$ (NGC\,4486 and NGC\,5846), do we reach a radius of $\sim R_e / 3$ only.

All data reduction was performed using the dedicated \XSauron\
software wrapped in a scripted pipeline (Paper~II).
For each target, individual datacubes were merged and analysed as
described in Paper~III, ensuring a minimum signal-to-noise ratio of 60 per pixel
using the binning scheme developed by \cite{CC03}. 
The \Sauron\ stellar kinematics were derived using a penalised pixel fitting routine 
\citep{CE04}, which provides parametric estimates of the 
line-of-sight velocity distribution (hereafter LOSVD) for each spaxel.
In Paper~III, we have presented the corresponding maps, which
include the mean velocity $V$, the velocity dispersion $\sigma$ and 
the Gauss-Hermite moments $h_3$ and $h_4$, for the 48 E and S0 \Sauron\ galaxies.

As mentioned in Paper~III, these quantities were measured fitting all $V$,
$\sigma$, $h_3$ and $h_4$ simultaneously: this ensures an optimal representation
of the corresponding LOSVD. In the present paper, we focus on the first two
true velocity moments, $\mu_1$ and $\mu_2$, which are sometimes
estimated by use of the Gauss-Hermite expansion of the LOSVD \citep{vdMF93}. As
emphasised in Paper~IV, the second order velocity moment 
is very sensitive to the details of the high velocity wings, which can rarely be 
accurately measured. We therefore decided to rely on a simpler but
more robust single gaussian fit (excluding higher order velocity moments),
and used the gaussian mean $V$ and standard deviation $\sigma$ to approximate the
first and second velocity moments. 

\subsection{Kinemetry}
\label{sec:kinemetry}

Following \cite{Krajnovic+06} who recently advocated 
the use of a method generalising the isophotal-shape tools 
\citep{Lauer85, Jedr87, BM87}, we employ kinemetry
as a quantitative approach to analyse the \Sauron\ stellar kinematic maps. 
Applying kinemetry on a velocity map provides radial profiles for
the kinematic position angle PA$_{kin}$, axis ratio $q_{kin}$, and
Fourier kinemetry terms, the dominant term $k_1$ representing the velocity amplitude.
We define a kinematic component to have constant or slowly 
varying PA$_{kin}$ and $q_{kin}$ radial profiles (taking into account the derived error bars). 
We then identify two separate components when we observe an abrupt change either with
$\Delta q_k > 0.1$, or $\Delta$PA$_{kin} > 10\degr$, 
or a double-hump in $k_1$ with a local minimum in between.
The transition between the two radial ranges is often emphasised by a peak
in the higher order $k_5$ Fourier term, 
which thus serves as an additional signature for such a change \citep{Krajnovic+06}.
Velocity maps which exhibit the presence of at least two stellar velocity components
are tagged as {\em Multiple Component} (MC), as opposed to {\em Single Component} (SC). 

We use a number of terms to represent some basic properties of the
individual kinematic components via quantitative criteria, following the definitions 
provided in \citet{Krajnovic+06} :
\begin{itemize}
\item {\em Low-level velocity} (LV): defined when the maximum velocity amplitude $k_1$ 
is lower than 15~\kms. Note that when the velocity amplitude is constant
over the field, PA$_{kin}$ and $q_{kin}$ are ill-defined.
When the central kinematical component is LV, we label the galaxy as
a central low-level velocity (CLV) system.
\item {\em Kinematic misalignment} (KM): defined when the absolute difference between 
the photometric and kinemetric position angles (PA$_{phot}$ and PA$_{kin}$) is larger than 10\degr.
\item {\em Kinematic twist} (KT): defined by a smooth variation 
of the kinematic position angle PA$_{kin}$ with an amplitude of at least 10\degr\
within the extent of the kinematic component.
\end{itemize}
\begin{figure*}
\centering
\epsfig{file=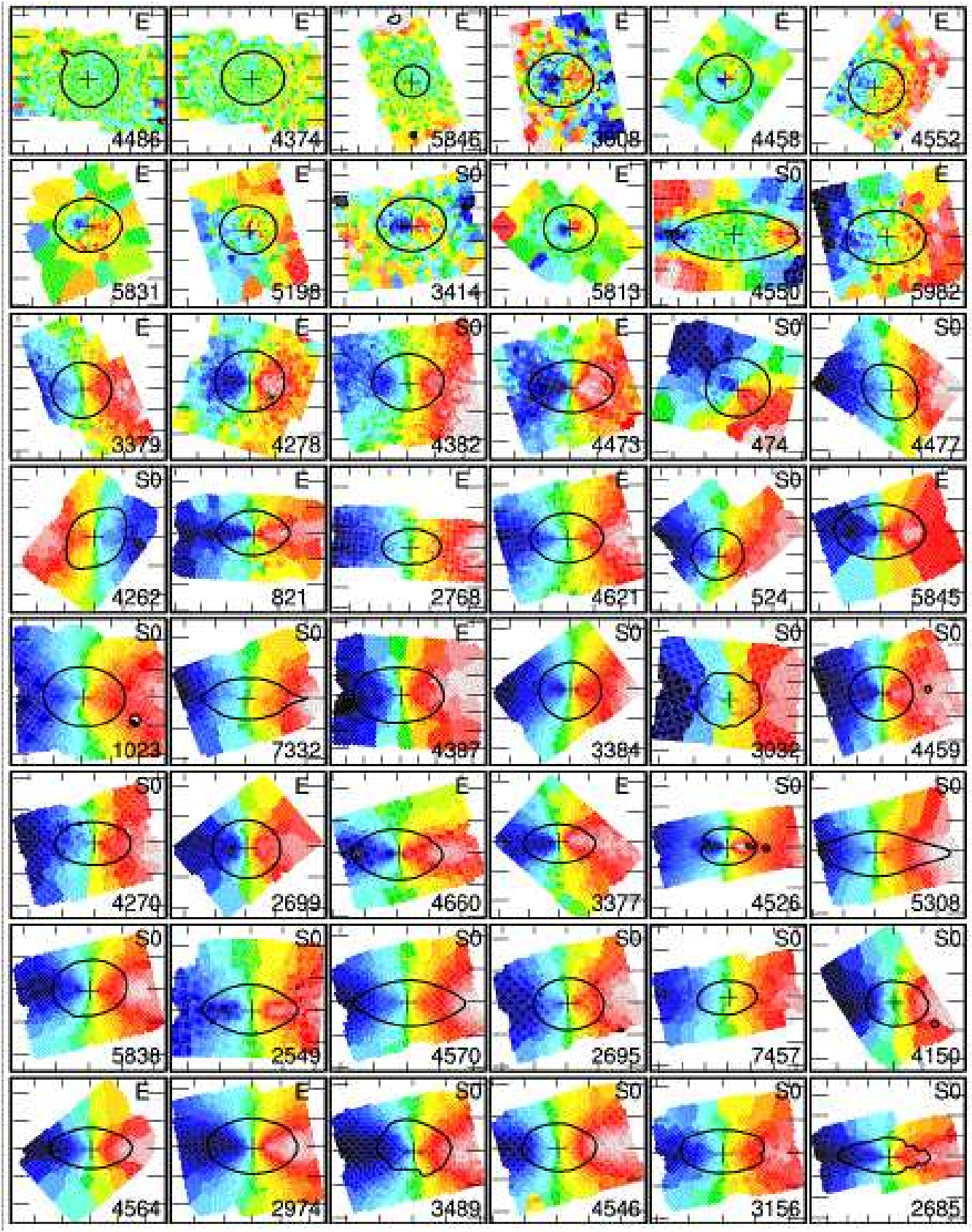, width=\hdsize}
\caption{\Sauron\ Stellar velocity fields for our 48 E and S0
galaxies (see Paper~III), the global outer photometric axis being horizontal. Colour cuts were tuned
for each individual galaxy as to properly emphasise the observed velocity structures. 
A representative isophote is overplotted in each thumbnail as a black solid line,
and the centre is marked with a cross.
Galaxies are ordered by increasing value of $\lambda_{R_e}$ (from left to right, top
to bottom; see Sect.~\ref{sec:LambdaR}). Slow rotators are galaxies on the first two rows.
NGC~numbers and Hubble types are provided in the lower-right and upper-right corners of each panel,
respectively. Tickmarks correspond to 10\arcsec.}
\label{fig:allV}
\end{figure*}

A {\em kinematically decoupled component} (KDC) is defined as an MC having
either an abrupt change in PA$_{kin}$, with a difference larger than 20\degr\ between 
two adjacent components, or an outer LV kinematic component (which prevents that 
component to have a robust PA$_{kin}$ measurement). This definition
roughly corresponds to the more standard appellation of
KDC used in the past \citep{Bender88}, although the two-dimensional
coverage provided by integral-field spectrographs allows a more sensitive detection procedure. 
These kinemetric groups will be examined in Sect.~\ref{sec:kinresults}.


\section{Quantifying the angular momentum}
\label{sec:LambdaR}

The velocity fields presented in Paper~III revealed a wealth of structures 
such as decoupled cores, velocity twists, misalignments, cylindrical 
or disc-like rotation (see also Section~\ref{sec:kinresults}).
It is difficult to disentangle the relative contributions of a true 
variation in the internal dynamical state and the effect of projection. 
A first examination of the stellar velocity maps for the 48 \Sauron\ E/S0 galaxies
(Fig.~\ref{fig:allV}, where velocity cuts have been adjusted as to properly emphasise the observed velocity structures) suggests that early-type galaxies come in two broad flavours: 
one which exhibits a clear large-scale and rather regular rotation pattern, 
and another which shows perturbed velocity structures (e.g. strong velocity twists) 
or central kinematically decoupled components with little rotation in the outer regions. 
One way to constrain the internal dynamics for a specific galaxy is to build 
a detailed model using all available observables. This was successfully achieved 
in Paper~IV on a subsample of 24 galaxies, for which accurate distances and 
high spatial resolution photometry are available, and where no strong signature 
of non-axisymmetry is observed. From these models, a considerable amount of 
detail on the orbital structure was derived (see also Paper~X).
We, however, also need a simple measurable parameter which quantifies the 
(apparent) {\em global} dynamical state of a galaxy, and which is applicable to
all galaxies in our representative sample of Es and S0s: this is the purpose
of this section.

\subsection{From $V/\sigma$ to $\lambda_R$ : a new kinematic parameter}
\label{sec:newkin}

The relation of $V / \sigma$ versus mean ellipticity $\epsilon$ 
(hereafter the anisotropy diagram) was often
used in the past to confront the apparent flattening with the observed amount of rotation
\citep{Illingworth77,Binney78, Davies+83}.
In the now classical treatment of the anisotropy diagram, the maximum observed 
rotational velocity $V_{\rm max}$ and the central dispersion $\sigma_0$ 
are generally used as surrogates for 
the mass-weighted mean of the square rotation speed and the random velocity.
This was mostly constrained by the fact that stellar kinematics were available 
along at most a few axes via long-slit spectroscopy.
\cite{Binney05} has recently revisited this diagram to design a more robust 
diagnostic of the velocity anisotropy in galaxies using two-dimensional
kinematic information. Starting from the Tensor Virial Theorem, 
Binney reformulated the ratio of ordered versus random motions 
in terms of integrated quantities observable with integral-field
spectrographs such as \Sauron, namely $\langle {V^2}\rangle$ and 
$\langle {\sigma^2}\rangle$, where ${V}$ and ${\sigma}$ 
denote respectively the observed stellar velocity and velocity dispersion, and the brackets 
correspond to a sky averaging weighted by the surface brightness (see Eq.~\ref{eq:brackets}). 

The \Sauron\ data provide us with a unique opportunity to derive for the first
time a robust measurement of $V/\sigma$ for a sample of local early-type galaxies.
We have therefore derived $\langle V^2\rangle$ and $\langle \sigma^2\rangle$
up to $\sim 1 R_e$, the resulting $V/\sigma$ values for the 48 \Sauron\ E and S0 galaxies being given 
in Table~\ref{tab:param}. S0s and Es tend to have on average relatively high and low $V/\sigma$, respectively.
A very significant overlap of the two types still exists, S0s reaching
values of $V/\sigma$ as low as 0.1, and Es as high as 0.7: as expected, the Hubble type 
can obviously not serve as a proxy for the galaxy kinematics (see the distribution
of Hubble types in Fig.~\ref{fig:allV}).

The availability of two-dimensional stellar kinematics 
makes $V/\sigma$ a useful tool to examine the dynamical status of early-type 
galaxies (see Paper~X). It fails, however, to provide us with a way to differentiate 
mean stellar velocity structures as different as those of 
NGC~3379 and NGC~5813 (see Fig.~\ref{fig:allV}). 
These two galaxies both have $V/\sigma \sim 0.14$,
and ellipticities $\epsilon$ in the same range (0.08 and 0.15, respectively), 
but their stellar velocity fields are qualitatively and quantitatively very
different: NGC~3379 displays a regular and large-scale rotation pattern
with a maximum amplitude of about 60~\kms, whereas NGC\,5813 exhibits 
clear central KDC with a peak velocity amplitude of $\sim 85$~\kms\ and 
a mean stellar velocity consistent with zero outside a radius of $\sim 12\arcsec$. 
The derivation of $V/\sigma$ includes a luminosity weighting that
amplifies the presence of the KDC in NGC\,5813. As a consequence,
NGC\,3379 with its global rotation and NGC\,5813 with its spatially confined
non-zero velocities end up with a similar $V/\sigma$.

We therefore need to design a new practical way to quantify the 
global velocity structure of galaxies using the two-dimensional spatial information
provided by integral-field units. The ideal tool would be a physical
parameter which captures the spatial information included in the kinematic
maps. Since we wish to assess the level of rotation in galaxies, 
this parameter should follow the nature of the classic $V / \sigma$: ordered
versus random motion. A measure of the averaged angular 
momentum $L = \langle \mathbf{R} \wedge \mathbf{V}\rangle $, could play the role of
$V$, and should be able to discriminate between large-scale rotation 
(NGC\,3379) and little or no rotation (NGC\,5813). Such a quantity, however, depends 
on the determination of the angular momentum vector direction, which is not 
an easily measured quantity. We therefore use a more robust and measurable quantity 
$\langle R \, \left| V \right| \rangle$, where $R$ is the observed distance to the
galactic centre, as a surrogate for $L$, and the brackets $\langle\,\rangle$ correspond
to a luminosity weighted sky average (see Eq.~\ref{eq:brackets}). We can then naturally define  
a dimensionless parameter, after normalising by e.g. mass, which leads to 
a proxy for the {\em baryon} projected specific angular momentum as:
\begin{equation}
\lambda_R \equiv \frac{\langle R \, \left| V \right| \rangle }{\langle R \, \sqrt{V^2 + \sigma^2} \rangle}
\, ,
\end{equation} 
and measured via two-dimensional spectroscopy as:
\begin{equation}
\label{eq:sumLambda}
\lambda_R = \frac{\sum_{i=1}^{N_p} F_i R_i \left| V_i \right|}{\sum_{i=1}^{N_p} F_i R_i \sqrt{V_i^2+\sigma_i^2}} \, ,
\end{equation}
where $F_i$ is the flux inside the $i^{\mbox{\small th}}$ bin, $R_i$ its distance to the centre,
and $V_i$ and $\sigma_i$ the corresponding mean stellar velocity and velocity dispersion.
The normalisation by the second velocity moment $V^2 + \sigma^2$ implies that
$\lambda_R$ goes to unity when the mean stellar rotation ($V$) dominates, with
$V^2 + \sigma^2$ being a reasonable proxy for mass (see Appendix~\ref{App:lambda}).
The use of higher order moments of either $V$ or the spatial weighting
$R$ would make this parameter more strongly dependent on the aperture and presence of
noise in the data. $\lambda_R$ obviously depends on the spatial extent over which
the sums in Eq.~\ref{eq:sumLambda} are achieved. In practice, we measure
$\lambda_R(R_m)$ within regions defined by the photometric best fit ellipses, where $R_m$ is the mean
radius of that ellipse ($a \sqrt{1 - \epsilon}$, with $a$ its semi-major axis and $\epsilon$ its ellipticity),
the area $A_{ellipse}$ of the corresponding aperture being thus $\pi R_m^2$, 
the area of a circle with a radius of $R_m$. 
When our \Sauron\ kinematic measurements do not fully sample the defined ellipse, 
we instead set the radius $R_m \equiv \sqrt{A_S / \pi}$ as that of a circular aperture with the same
area $A_{S}$ on the sky actually covered by the \Sauron\ data inside that aperture (see Paper~IV).
For a specific galaxy, we measure $\lambda_R(R_m)$ up to the radius for which we reach a 
maximum difference of 15\% between $A_{S}$ and $A_{ellipse}$: this guarantees that the \Sauron\ 
kinematic data still properly fill up the elliptic aperture defined by the photometry.
\begin{figure}
\centering
\epsfig{file=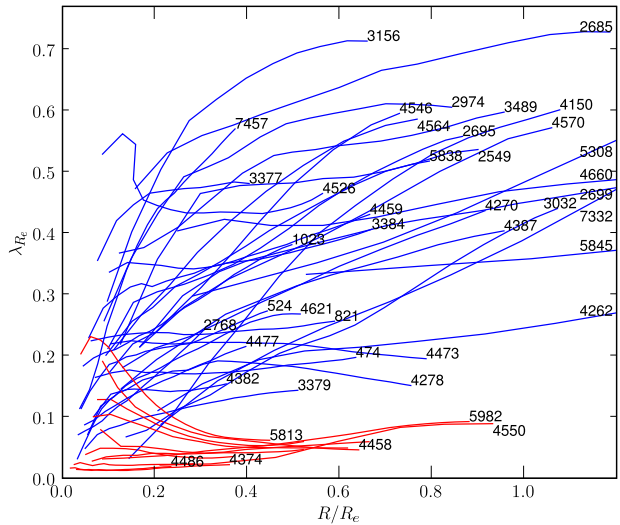, width=\columnwidth}
\caption{Radial $\lambda_R$ profiles for the 48 E and S0 galaxies of the \Sauron\ sample. 
Profiles of slow and fast rotators are coloured in red and blue, respectively. NGC~numbers are
indicated for all fast rotators and most slow rotators (a few were removed for legibility).}
\label{fig:profLR}
\end{figure}

\subsection{Rotators and $\lambda_R$}
\label{sec:rotators}

Values of $\lambda_{R_e}$, $\lambda_R$ within 1~$R_e$ or the largest radius allowed by our \Sauron\ data, 
whichever is smaller, are provided in Table~\ref{tab:param} for the 48 E/S0 galaxies.
As expected, NGC\,3379 and NGC\,5813 which 
have similar $V/\sigma$ values (see Fig.~\ref{fig:allV} 
and Sect.~\ref{sec:newkin}) but {\em qualitatively} different 
velocity structures, are now {\em quantitatively} distinguished with 
$\lambda_{R_e}$ values of 0.14 and 0.06, respectively.
This difference is significant because the formal uncertainty in 
the derivation of $\lambda_{R_e}$ due to the presence of noise in the data is 
almost always negligible, and anyway smaller than 0.02 for galaxies such 
as NGC\,3379 and NGC\,5813 (Appendix \ref{App:2I}): this can be 
understood because $\lambda_{R_e}$ includes averages over a large area.
There is, however, a systematic (positive) bias which obviously increases as the 
the velocity amplitudes in the galaxy decrease, and can reach up
to about $0.03$ in the measurement of $\lambda_{R_e}$. This bias is 
therefore dominant for the three galaxies with very low $\lambda_{R_e}$ ($< 0.03$), 
the mean stellar velocities being in fact consistent with 
zero values everywhere in the field of view. 

As we go from galaxies with low to high $\lambda_{R_e}$ values, the overall velocity
amplitude naturally tends to increase. More importantly, there seems to be 
a {\em qualitative} change in the observed stellar velocity structures.
This is already illustrated in Fig.~\ref{fig:allV}, where the 48
\Sauron\ stellar velocity fields are ordered, from left to right, top to bottom, by
increasing value of $\lambda_{R_e}$. Rotators with $\lambda_{R_e} <0.1$ exhibit
low stellar mean velocities at large radii, with very perturbed stellar kinematics and 
large-scale kinematically decoupled components (this point
will be further examined in Sect.~\ref{sec:kinresults}).

\begin{figure}
\centering
\epsfig{file=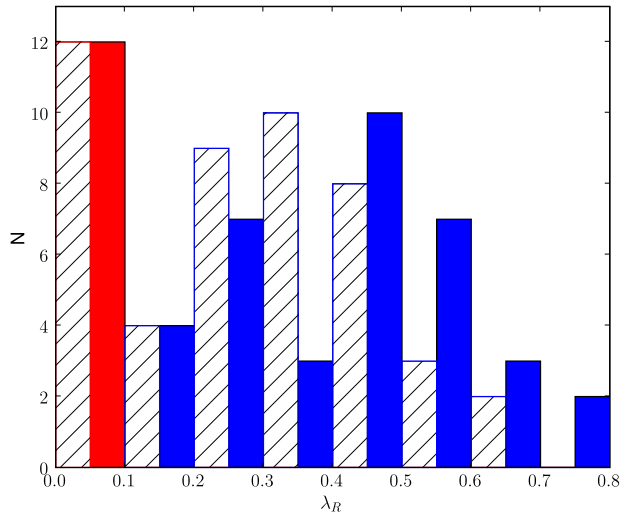, width=\columnwidth}
\caption{Histogram of $\lambda_{R}$: bars with plain colours correspond to $\lambda_{R_e}
\equiv \lambda_{R}(R_e)$ and the dashed bars to $\lambda_{R}(R_e/2)$. 
Red bars indicate the $\lambda_{R}$ bin for slow rotators:
the number of slow rotators does not change if we were to measure $\lambda_{R}$ at $R_e/2$.}
\label{fig:hist}
\end{figure}
This qualitative change is nicely illustrated in Fig.~\ref{fig:profLR} where we show
the radial $\lambda_R$ profiles. Galaxies with $\lambda_{R_e}$ below and above 0.1
exhibit qualitatively very different $\lambda_R$ profiles: the former
have either decreasing or nearly flat (and small amplitude) $\lambda_R$ profiles, 
while the latter preferentially exhibit significantly increasing $\lambda_R$ radial profiles.
Observed $\lambda_R$ gradients are therefore negative or rather small within 1~$R_e$ for galaxies 
with $\lambda_{R_e} < 0.1$, and we will label them as ``slow rotators".
This contrasts with the significantly rising $\lambda_R$ profiles for galaxies 
with $\lambda_{R_e} > 0.1$, which all exhibit clear large-scale and 
relatively regular rotation patterns, and which we label, by opposition, as ``fast rotators":
the fact that this class includes both mild and very fast rotators
is discussed in Sect.~\ref{sec:LRVS}.

As mentioned, Table~\ref{tab:param} includes $\lambda_{R_e}$ values derived 
using the \Sauron\ two-dimensional kinematic maps available and a default equivalent aperture of
1~$R_e$. This aperture is in fact covered by the \Sauron\ datacubes 
for 17 galaxies out of the 48 in the \Sauron\ E/S0 sample (see Paper IV), 
with two galaxies being mapped only to $\sim 0.3~R_e$ (NGC\,4486 and NGC\,5846). 
Galaxies with the narrowest relative spatial coverage
(with the largest $R_e$: NGC\,4486, NGC\,5846) are among the slowest rotators of our sample: 
these two galaxies are in fact known not to exhibit any significant rotation within 1~$R_e$ 
\citep[see][]{Sembach96+}, even though NGC\,4486 (M\,87) is in fact flattened 
at very large radii \citep[see][and references therein]{Kissler98+}. 
This implies that only a few galaxies near the $\lambda_{R_e} = 0.1$ threshold 
(NGC\,5982, NGC\,4550, NGC\,4278) could cross that threshold
if we were to have a complete coverage up to 1~$R_e$. 

This is also illustrated in Fig.~\ref{fig:hist} which shows that histograms of $\lambda_{R_e}$ for
radii of 1~$R_e$ and $R_e/2$ (or restricted to the equivalent effective aperture 
of the \Sauron\ field of view, whichever is smaller) have the same fraction of slow and fast rotators.
The overall distribution of $\lambda_R$ values is similar for $R_e$ and $R_e/2$, although
$\lambda_R$ tends to increase significantly from $R_e / 2$ to $R_e$ for fast rotators (see also the solid
straight lines in Fig.~\ref{fig:RVS}), 
an obvious implication of the observed rising $\lambda_R$ profiles in Fig.~\ref{fig:profLR}. 
There are 36 fast rotators and 12 slow rotators (75\% and 25\% of the total sample), 
their median $\lambda_{R_e}$ values being $\sim 0.44$ and $0.05$ respectively.
Within the class of slow rotators, three galaxies have $\lambda_{R_e}$ significantly 
below 0.03 (their mean stellar velocity maps being consistent with 
zero rotation everywhere, as mentioned above). These are among the brightest
galaxies of our sample, namely NGC\,4486, NGC\,4374 and NGC\,5846.

\subsection{Misalignments and twists}
\label{sec:misalign}
\begin{figure}
\label{fig:kinemetry}
\centering
\epsfig{file=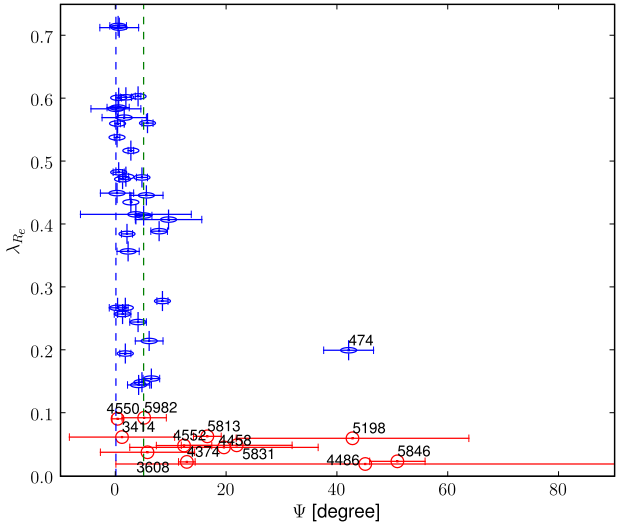, width=\columnwidth}
\caption{$\lambda_{R_e}$ versus the kinematic misalignment $\Psi$ between 
the global photometric major-axis and the kinematic axis within the \Sauron\ field. 
Slow rotators are represented by red circles, fast rotators by special blue symbols
(horizontal ellipse plus a vertical line). 
The vertical dashed line corresponds to $\Psi = 5\degr$.
Nearly all fast rotators have small $\Psi$ values ($< 10\degr$), the only exception
being NGC\,474, the photometry of which is perturbed by the presence of irregular shells. 
This contrast with slow rotators which show significantly non-zero $\Psi$ values.
}
\label{fig:misalign}
\end{figure}
The fact that slow and fast rotators exhibit distinct kinematics can be demonstrated by considering
the global alignment (or misalignment) between photometry and kinemetry.
Fig.~\ref{fig:misalign} illustrates this by showing the kinematic misalignment $\Psi \equiv
| \mbox{PA}_{phot} - \mbox{PA}_{kin}|$ for all
48 E and S0 galaxies in the \Sauron\ sample. Note that the photometric PA is
derived using the large-scale MDM data, but the kinematic PA via a global measurement on 
the \Sauron\ velocity maps as described in Appendix~C of \cite{Krajnovic+06}.
All fast rotators, except one, have misalignments $\Psi$ below 10\degr.
The only exception is NGC\,474 which is an interacting galaxy with well-know irregular shells 
\citep{Turnbull+99}.  In fact, the few galaxies which have $5\degr < \Psi < 10\degr$ 
(NGC\,3377, NGC\,3384, NGC\,4382, NGC\,4477, NGC\,7332) are almost certainly barred.
Even the two relatively face-on galaxies NGC\,4262 and NGC\,4477, the photometry of which
shows the signature of a strong bar within the \Sauron\ field of view, have their {\em outer} photometric PA 
well aligned with the measured kinematic PA.
In contrast, more than half of all slow rotators have $\Psi > 10\degr$, and
none of these exhibit any hint of a bar. This difference in the misalignment values
of slow and fast rotators cannot be entirely due to the effect of inclination, not only because of
the argument mentioned in Sect.~\ref{sec:KB96}, but also because even the roundest fast rotators
do not exhibit large misalignment values (see Sect.~5.1 in Paper~X).

Another remarkable feature comes from the observed velocity twists in the \Sauron\ kinematic maps:
only 6 galaxies out of 48 exhibit strong velocity twists larger than 30\degr\ outside the 
inner 3\arcsec, namely NGC\,3414, 3608, 4550, 4552, 5198, and 5982, with 3 out of these
6 having large-scale counter-rotating stellar components (NGC\,3414, 3608, and 4550). 
All these galaxies are in fact slow rotators. 
This implies that only fast rotators have a relatively well defined
(apparent) kinematic major-axis, which in addition is
roughly aligned with the photometric major-axis.

There is a close link between $\lambda_R$ and the (apparent) specific angular momentum 
(see Appendix~\ref{App:lambda}). $\lambda_R$ is therefore a continuous parameter
which provides a {\em quantitative} assessment of the apparent mean stellar rotation.
As emphasised in this Section, slow and fast rotators exhibit quantitatively 
but also qualitatively different stellar kinematics. This strongly suggests
that slow rotators as a class cannot simply be scaled-down versions of 
galaxies with $\lambda_{R_e} > 0.1$.

\subsection{Inclination effects}
\label{sec:LRVS}

Obviously, $\lambda_R$ is derived from projected quantities and therefore
also significantly depends on the viewing angles.
Can slow rotators be face-on versions of fast rotators?
Assuming $\lambda_R$ roughly follows the behaviour of $V/\sigma$ with the inclination
angle $i$ (Appendix~\ref{App:2I}), a galaxy with a {\em measured} $\lambda_{R_e} = 0.05$, 
typical of observed slow rotators, would require to be at a nearly face-on 
inclination of $i \sim  20\degr$ to reach an {\em intrinsic} (edge-on) value of $\lambda_{R_e} \sim 0.15$, 
the smallest value for all observed fast rotators in our sample. 
The probability that slow rotators are truly fast rotators seen face-on 
is therefore small, and cannot explain the 25\% population of slow rotators observed in the
$\lambda_{R_e}$ histogram (Fig.~\ref{fig:hist}). 

As emphasised above, the measured $\lambda_{R_e}$ values 
are most probably lower limits because we observe galaxies away from edge-on.
An illustrative example is provided by the case of NGC\,524. 
According to Paper~IV, NGC\,524 is viewed with an inclination of 19\degr: 
still, NGC\,524 exhibits a regular velocity pattern, and is a 
fast rotator with $\lambda_{R_e} \sim 0.28$. 
There is also some evidence that the three fast rotators
with the lowest $\lambda_{R_e}$ values in our sample, namely NGC\,3379, NGC\,4278 and 
NGC\,4382, are significantly inclined galaxies (\citet{Statler01}; 
Papers~III, IV and \citet{PaperV}, hereafter Paper~V; \citet{Fisher97}). 
This illustrates the fact that the apparent distribution
of $\lambda_{R_e}$ {\em within} the fast rotator class may not fully reflect the 
intrinsic distribution with galaxies viewed edge-on.
It is, however, impossible to estimate the ``edge-on" $\lambda_{R_e}$
distribution for fast rotators as this would require {\em ad minima} an accurate measurement
of their inclination. Although the fast rotator class may probably 
include truly mild as well as very fast rotators
(as expected from the continuous nature of $\lambda_{R_e}$),
all galaxies with $\lambda_{R_e}> 0.1$ exhibit a global and regular rotation pattern, with a clear
sense of rotation and a significant amount of specific stellar angular momentum. This, to us,
justifies the use of the label ``fast rotators" for these galaxies. Further
discussions of the true nature of fast rotators as a class 
has to wait for a larger and complete sample.
\begin{table*}
\begin{minipage}{177mm}
\caption{Characteristics of the E and S0 galaxies in the representative \Sauron\ sample. All galaxies with 
$\lambda_{R_e} > 0.1$ are classified as fast rotators.}
\label{tab:param}
\begin{tabular}{rlcccccccccccrc}
\hline
Galaxy & Type & $T$ & $(m-M)$ & $M_B$ & $R_{\rm e}$ & $\sigma'_e$ & $\epsilon_{1/2}$ & $\epsilon$ & $\epsilon_e$ & $(a_4/a)_e$ & $V/\sigma$ & $\lambda_{R_e}$ & Group & Rotator \\
(NGC)  &       &     & (mag) & (mag) & (arcsec) & (km/s) &     &     &      &      &      &      &      &  \\
 (1) & ~~(2) & (3) & (4)   & (5)   & (6)      & ~(7) & (8) & (9) & (10) & (11) & (12) & (13) & (14) & (15)\\
\hline
\noalign{\smallskip}
474  & S0$^0$(s)               & -2.0& 32.56 &  -20.42 &  29 &  150  &  0.11 & 0.11 & 0.13 & -0.14 &  0.21 &  0.200  &  MC   & F \\    
524  & S0$^+$(rs)              & -1.2& 31.84 &  -21.40 &  51 &  235  &  0.05 & 0.05 & 0.04 & -0.16 &  0.29 &  0.278  &  SC   & F \\    
821  & E6?                     & -4.8& 31.85 &  -20.44 &  39 &  189  &  0.40 & 0.40 & 0.35 &  1.43 &  0.26 &  0.258  &  SC   & F \\  
1023 & SB0$^-$(rs)             & -2.7& 30.23 &  -20.42 &  48 &  182  &  0.33 & 0.33 & 0.36 &  0.54 &  0.35 &  0.385  &  SC   & F \\    
2549 & S0$^0$(r)~sp            & -2.0& 30.44 &  -19.36 &  20 &  145  &  0.47 & 0.49 & 0.49 &  2.86 &  0.56 &  0.539  &  MC   & F \\    
2685 & (R)SB0$^+$pec$\!\!\!\!$ & -1.1& 31.15 &  -19.05 &  20 &   96  &  0.56 & 0.62 & 0.59 &  2.93 &  0.88 &  0.716  &  SC   & F \\    
2695 & SAB0$^0$(s)             & -2.1& 32.49 &  -19.38 &  21 &  188  &  0.27 & 0.29 & 0.21 &  0.36 &  0.54 &  0.561  &  MC   & F \\    
2699 & E:                      & -5.0& 32.09 &  -18.85 &  14 &  124  &  0.16 & 0.15 & 0.19 &  1.04 &  0.43 &  0.450  &  MC   & F \\  
2768 & E6:                     & -4.3& 31.69 &  -21.15 &  71 &  216  &  0.38 & 0.38 & 0.46 &  0.12 &  0.24 &  0.268  &  SC   & F \\  
2974 & E4                      & -4.7& 31.60 &  -20.32 &  24 &  233  &  0.38 & 0.37 & 0.37 &  0.64 &  0.70 &  0.602  &  SC   & F \\  
3032 & SAB0$^0$(r)             & -1.8& 31.65 &  -18.77 &  17 &   90  &  0.15 & 0.11 & 0.17 &  0.44 &  0.27 &  0.416  &  CLV  & F \\    
3156 & S0:                     & -2.4& 31.69 &  -18.08 &  25 &   65  &  0.49 & 0.47 & 0.47 & -0.04 &  0.88 &  0.713  &  SC   & F \\    
3377 & E5-6                    & -4.8& 30.19 &  -19.24 &  38 &  138  &  0.46 & 0.46 & 0.50 &  0.94 &  0.49 &  0.475  &  SC   & F \\  
3379 & E1                      & -4.8& 30.06 &  -20.16 &  42 &  201  &  0.08 & 0.08 & 0.11 &  0.16 &  0.14 &  0.145  &  SC   & F \\  
3384 & SB0$^-$(s):             & -2.7& 30.27 &  -19.56 &  27 &  145  &  0.20 & 0.20 & 0.20 &  1.13 &  0.44 &  0.414  &  MC   & F \\  
3414 & S0~pec                  & -2.1& 31.95 &  -19.78 &  33 &  205  &  0.21 & 0.21 & 0.23 &  1.80 &  0.09 &  0.062  &  KDC  & S \\    
3489 & SAB0$^+$(rs)$\!\!$      & -1.3& 30.35 &  -19.32 &  19 &   98  &  0.28 & 0.29 & 0.29 & -0.61 &  0.67 &  0.602  &  MC   & F \\  
3608 & E2                      & -4.8& 31.74 &  -19.54 &  41 &  178  &  0.18 & 0.18 & 0.20 & -0.21 &  0.05 &  0.038  &  KDC  & S \\  
4150 & S0$^0$(r)?              & -2.1& 30.64 &  -18.48 &  15 &   77  &  0.25 & 0.30 & 0.28 & -0.32 &  0.58 &  0.584  &  CLV  & F \\  
4262 & SB0$^-$(s)              & -2.7& 31.23 &  -18.88 &  10 &  172  &  0.08 & 0.22 & 0.11 &  1.28 &  0.24 &  0.245  &  MC   & F \\  
4270 & S0                      & -1.9& 32.83 &  -18.28 &  18 &  122  &  0.41 & 0.50 & 0.44 & -0.64 &  0.40 &  0.446  &  MC   & F \\  
4278 & E1-2                    & -4.8& 30.97 &  -19.93 &  32 &  231  &  0.12 & 0.12 & 0.13 & -0.15 &  0.18 &  0.149  &  MC   & F \\  
4374 & E1                      & -4.2& 31.27 &  -21.23 &  71 &  278  &  0.15 & 0.15 & 0.13 & -0.40 &  0.03 &  0.023  &  SC   & S \\  
4382 & S0$^+$(s)pec$\!$        & -1.3& 31.27 &  -21.28 &  67 &  196  &  0.19 & 0.19 & 0.22 &  0.59 &  0.16 &  0.155  &  CLV  & F \\  
4387 & E                       & -4.8& 31.59 &  -18.34 &  17 &   98  &  0.35 & 0.40 & 0.32 & -0.76 &  0.39 &  0.408  &  SC   & F \\  
4458 & E0-1                    & -4.8& 31.12 &  -18.42 &  27 &   85  &  0.12 & 0.12 & 0.14 &  0.41 &  0.12 &  0.046  &  KDC  & S \\  
4459 & S0$^+$(r)               & -1.4& 30.98 &  -19.99 &  38 &  168  &  0.18 & 0.17 & 0.17 &  0.22 &  0.45 &  0.436  &  MC   & F \\  
4473 & E5                      & -4.7& 30.92 &  -20.26 &  27 &  192  &  0.39 & 0.41 & 0.43 &  1.03 &  0.22 &  0.195  &  MC   & F \\  
4477 & SB0(s):?                & -1.9& 31.07 &  -19.96 &  47 &  162  &  0.24 & 0.24 & 0.23 &  2.04 &  0.21 &  0.215  &  SC   & F \\  
4486 & E0-1$^+$pec             & -4.3& 30.97 &  -21.79 & 105 &  298  &  0.04 & 0.04 & 0.07 & -0.07 &  0.02 &  0.019  &  SC   & S \\  
4526 & SAB0$^0$(s):$\!$        & -1.9& 31.08 &  -20.68 &  40 &  222  &  0.36 & 0.37 & 0.41 & -1.92 &  0.54 &  0.476  &  MC   & F \\  
4546 & SB0$^-$(s):             & -2.7& 30.69 &  -19.98 &  22 &  194  &  0.39 & 0.45 & 0.36 &  0.69 &  0.60 &  0.604  &  MC   & F \\  
4550 & SB0$^0$:sp              & -2.0& 30.94 &  -18.83 &  14 &  110  &  0.58 & 0.61 & 0.62 &  2.36 &  0.10 &  0.091  &  MC   & S \\  
4552 & E0-1                    & -4.6& 30.86 &  -20.58 &  32 &  252  &  0.04 & 0.04 & 0.06 &  0.00 &  0.05 &  0.049  &  KDC  & S \\  
4564 & E                       & -4.8& 30.82 &  -19.39 &  21 &  155  &  0.47 & 0.52 & 0.43 &  1.33 &  0.58 &  0.586  &  SC   & F \\  
4570 & S0~sp                   & -2.0& 31.23 &  -19.54 &  14 &  173  &  0.41 & 0.60 & 0.44 &  1.90 &  0.53 &  0.561  &  MC   & F \\  
4621 & E5                      & -4.8& 31.25 &  -20.64 &  46 &  211  &  0.34 & 0.34 & 0.35 &  1.66 &  0.25 &  0.268  &  KDC  & F \\  
4660 & E                       & -4.7& 30.48 &  -19.22 &  11 &  185  &  0.33 & 0.44 & 0.41 &  0.66 &  0.49 &  0.472  &  MC   & F \\  
5198 & E1-2:                   & -4.7& 33.06 &  -20.38 &  25 &  179  &  0.14 & 0.12 & 0.14 & -0.17 &  0.07 &  0.060  &  KDC  & S \\  
5308 & S0$^-$~sp               & -2.0& 32.65 &  -20.27 &  10 &  208  &  0.32 & 0.60 & 0.53 &  4.74 &  0.45 &  0.483  &  MC   & F \\    
5813 & E1-2                    & -4.8& 32.38 &  -20.99 &  52 &  230  &  0.15 & 0.15 & 0.17 & -0.03 &  0.14 &  0.063  &  KDC  & S \\  
5831 & E3                      & -4.8& 32.11 &  -19.73 &  35 &  151  &  0.15 & 0.15 & 0.20 &  0.46 &  0.08 &  0.049  &  KDC  & S \\  
5838 & S0$^-$                  & -2.7& 32.37 &  -19.87 &  23 &  240  &  0.27 & 0.34 & 0.28 &  0.34 &  0.51 &  0.518  &  MC   & F \\  
5845 & E:                      & -4.8& 32.01 &  -18.58 &   4 &  239  &  0.35 & 0.35 & 0.31 &  0.63 &  0.36 &  0.357  &  MC   & F \\  
5846 & E0-1                    & -4.7& 31.92 &  -21.24 &  81 &  238  &  0.07 & 0.07 & 0.07 & -0.38 &  0.03 &  0.024  &  SC   & S \\  
5982 & E3                      & -4.8& 33.15 &  -21.46 &  27 &  229  &  0.30 & 0.30 & 0.28 & -0.92 &  0.08 &  0.093  &  KDC  & S \\  
7332 & S0~pec~sp               & -2.0& 31.40 &  -19.93 &  11 &  125  &  0.40 & 0.42 & 0.39 &  1.35 &  0.32 &  0.390  &  KDC  & F \\    
7457 & S0$^-$(rs)?             & -2.6& 30.55 &  -18.81 &  65 &   78  &  0.44 & 0.44 & 0.43 &  0.20 &  0.62 &  0.570  &  CLV  & F \\    
\hline
\end{tabular}
\\
Notes:
(1)~Galaxy identifier (NGC number). 
(2)~Hubble type (NED). 
(3)~Numerical morphological type \citep[LEDA][]{Patu03}.
(4)~Galaxy distance modulus from \citet{Tonry+01,Tully88}, \citep[corrected by subtracting 0.06 mag, see][]{Mei+05}, or from the LEDA database assuming a Hubble flow with $H = 75$~km\,s$^{-1}$\,Mpc$^{-1}$.
(5)~Absolute $B$ magnitude (Paper~II).
(6)~Effective radius, in arcsec.
(7)~Velocity dispersion derived using the luminosity-weighted spectrum within $R_e$ or within the \Sauron\ field, whichever is smaller.
(8) and (9)~Global ellipticity within $R_e / 2$ and $R_e$ (from \Sauron), or within the \Sauron\ field, whichever is smaller.
(10)~Mean ellipticity within 1~$R_e$, derived from the GALPHOT radial profile.
(11)~Mean isophote shape parameter $a_4/a$ (in \%) within 1~$R_e$.
(12) and (13)~$V/\sigma$ (see Paper~X) and $\lambda_R$ within 1 $R_e$.
(14)~Kinemetry group (see Sect.~\ref{sec:kinemetry}).
(15)~Rotator class: F$=$fast, S$=$slow.
\end{minipage}
\end{table*}

\section{Other classifications}
\label{sec:Hubble}

In this Section, we examine in more detail other galaxy properties,
including photometry and kinemetry, to determine if these could help
in defining kinematic classes, and their potential link with $\lambda_R$.

\subsection{Hubble Classification}
\label{sec:ES0}

There is a clear overlap of Es and S0s in the fast rotator class, 
with most of the corresponding galaxies having ellipticities higher than 0.2. 
This illustrates the fact that many Es have kinematic characteristics similar to
S0s \citep{Bender88, RixWhite90}. Some galaxies classified as ellipticals 
also show photometric signatures of embedded discs, quantified as a positive $a_4 / a$,
and were classified as ``discy" ellipticals or E(d) by KB96.
It seems difficult, however, to distinguish between a so-called E(d) and an S0
galaxy from the photometric properties or the anisotropy diagram alone. 
KB96 argued that E(d)s are objects intermediate between S0s and 
boxy ellipticals or E(b)s (ellipticals with boxy
isophotes, i.e. negative $a_4 / a$). In principle, the Ed-S0-Sa sequence is
one of decreasing bulge-to-total light ratio ($B/T$), with S0s having $B / T
\sim 0.6$ in the B band \citep{FAlven04}. $B/T$ is, however, a
fairly difficult quantity to measure in early-type galaxies, as it depends on
the adopted model for the surface brightness distribution of the disc and
bulge components. As emphasised by \cite{deJong+04}, the kinematic
information is critical in assessing the rotational support of both components.

Consider two galaxies of our sample, NGC~3377
and NGC~2549, which both have similar $\lambda_{R_e}$, 
the former being classified as a discy elliptical, the latter as a lenticular. Both have
very similar total luminosity, $V-I$ colour \citep{Tonry+01} and gas content 
(Paper~V). Their \Sauron\ stellar kinematic maps are
also quite similar (Paper~III), with large-scale disc-like rotation,
a centrally peaked stellar velocity dispersion and a significant $h_3$ term,
anti-correlated with the mean velocity. Finally, there is evidence that 
NGC~3377 contains a bar, with its velocity map exhibiting a stellar kinematical
misalignment and a spiral-like ionised gas distribution (Paper~V). 
This therefore strongly argues for NGC~3377 to be a
misclassified barred S0 (SAB0). Many authors \citep[e.g.][and KB96]{vdB90, Michard94, JorgensenFranx94}
indeed suggested that many Es are in fact misclassified S0s, mostly 
on the basis of photometry alone. Considering the kinematic properties of the observed 
\Sauron\ early-type galaxies, we suspect that 
most and possibly all of the 13 Es which are fast rotators are in fact misclassified 
S0 galaxies, a conclusion also supported by the results of Paper~X via the use of 
state-of-the-art dynamical models.

\subsection{Isophote shapes}
\label{sec:isophotes}

\begin{figure}
\centering
\epsfig{file=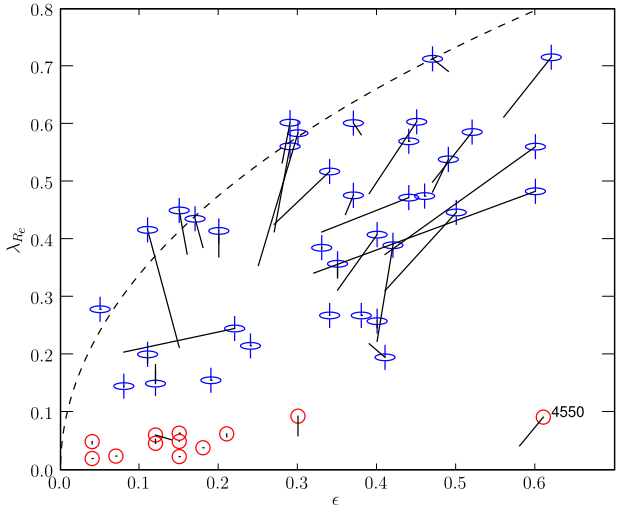, width=\columnwidth}
\caption{$\lambda_{R_e}$ versus the global ellipticity $\epsilon$ for the 48 E and
S0 galaxies of the \Sauron\ sample. The solid lines indicate where the points would move
for a smaller equivalent effective aperture of $R_e/2$. Slow rotators (which have 
$\lambda_{R_e} \leqslant 0.1$) are red circles, and fast rotators ($\lambda_{R_e} > 0.1$) are represented by
special blue symbols (horizontal ellipse plus a vertical line).
The black dashed line corresponds to the curve expected for isotropic oblate rotators
viewed edge-on \citep[see Appendix~\ref{App:2I} and][]{Binney05}. The galaxy NGC\,4550
is labelled.}
\label{fig:RVS}
\end{figure}
Figure~\ref{fig:RVS} shows $\lambda_{R_e}$ versus the global ellipticity $\epsilon$ for the
48 \Sauron\ E and S0s, measured using an aperture of 1~$R_e$ (or including 
the full \Sauron\ field of view for galaxies with large $R_e$, see Sect.~\ref{sec:rotators}). 
All galaxies in Fig.~\ref{fig:RVS} lie close to or 
below the curve expected for isotropic oblate rotators
viewed edge-on \citep[see Appendix~\ref{App:2I} and][]{Binney05}.
Fast rotators, which have velocity maps with significant large-scale rotation, 
have ellipticities ranging up to about 0.6. Apart from the atypical case of NGC~4550, 
slow rotators show little or spatially confined rotation
and all have ellipticities $\epsilon < 0.3$.
NGC\,4550 is in fact a nearly edge-on galaxy with
two co-spatial counter-rotating stellar discs, each contributing for about 50\% of the total
luminosity, and should therefore be regarded as an atypical case where a high ellipticity is
accompanied by a relatively low mean stellar velocity in the equatorial plane
\citep[][and see Paper~X for a detailed discussion]{Rix+92,Rubin+92}.

\label{sec:KB96}
\begin{figure}
\centering
\epsfig{file=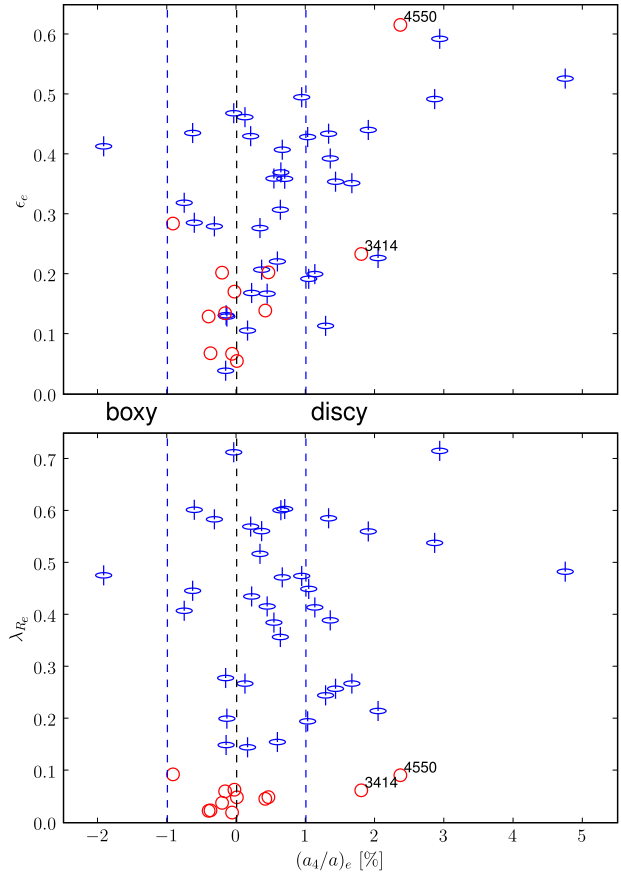, width=\columnwidth}
\caption{Mean ellipticity $\epsilon_e$ (top) and $\lambda_{R_e}$ (bottom)
versus mean $(a_4/a)_e$ (in \%) within 1~$R_e$. 
Symbols for slow and fast rotators are as in Fig.~\ref{fig:RVS}. The vertical dashed lines
 correspond to $(a_4/a)_e = \pm 1\%$}
\label{fig:eps_a4}
\end{figure}
\begin{figure*}
\centering
\epsfig{file=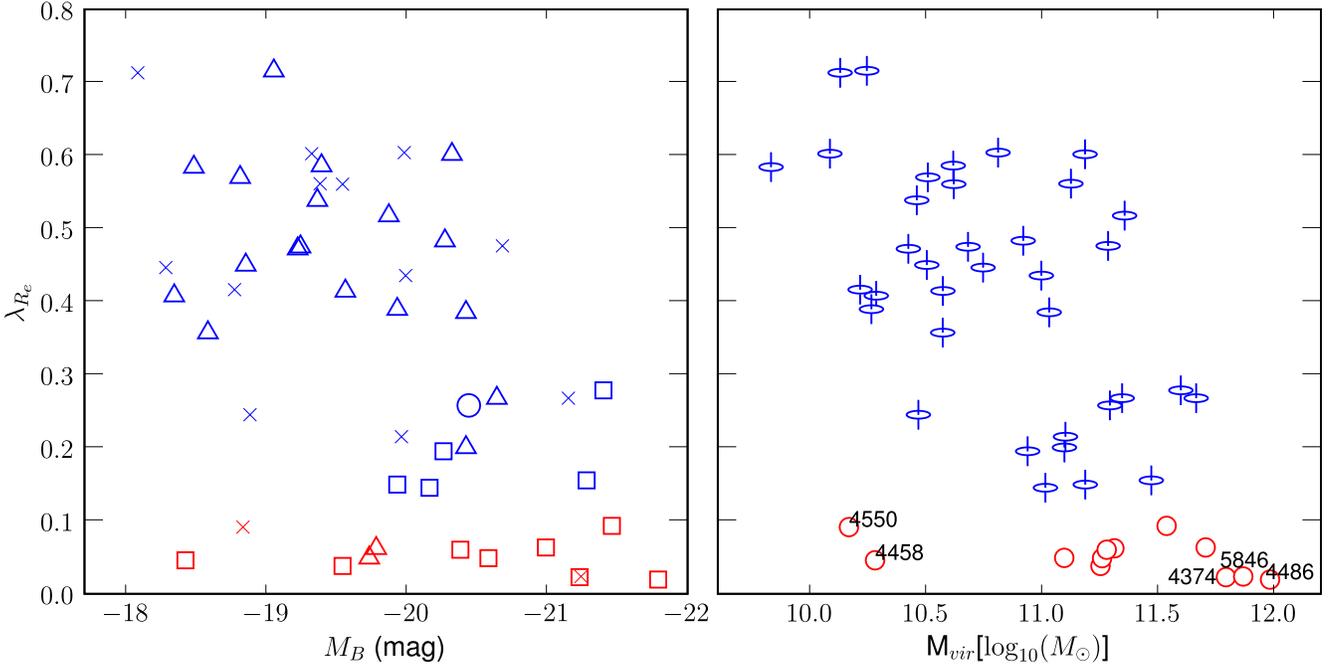, width=\hdsize}
\caption{$\lambda_{R_e}$ versus absolute magnitude $M_B$ (left panel) and virial mass
M$_{vir}$ (right panel) for the 48 E and S0 of the \Sauron\ sample. 
In the left panel, symbols correspond to the cusp slope 
classification \citep{Faber+97, Rest+01, Ravindranath+01, Lauer+05} with
power-laws as triangles, cores as squares, NGC\,821 which is an ``intermediate" object as a circle, and
crosses indicating galaxies for which there is no published classification.
In the right panel, symbols for slow and fast rotators are as in Fig.~\ref{fig:RVS}, and 
NGC numbers for a few galaxies are indicated.}
\label{fig:LRMb}
\end{figure*}
KB96 mention that $a_4/a$ can be used as a reasonably reliable way to measure
velocity anisotropy. They first observe, using $(V/\sigma)^{\star}$ ($V/\sigma$ 
normalised to the value expected for an isotropic edge-on oblate system), 
that discy galaxies ($a_4 / a > 0$) seem consistent with near isotropy, whereas boxy galaxies ($a_4 / a < 0$) spread over
a large range of $(V/\sigma)^{\star}$ values. However, as emphasised by \cite{Binney05} and
in Paper~X, $(V/\sigma)^{\star}$ is not a good indicator of anisotropy:
the relation between $(V/\sigma)^{\star}$ and the anisotropy of a galaxy
strongly depends on its flattening, as well as on its inclination \citep{BN05}.
More flattened galaxies will thus tend to lie closer to the $(V/\sigma)^{\star} = 1$ curve, which
explains in part why galaxies with discy and boxy isophotes seem to extend over
different ranges of $(V/\sigma)^{\star}$, the former being
then incorrectly interpreted as near-isotropic systems. 
This suggests we should avoid using $(V/\sigma)^{\star}$ at all,
and leads us to examine more directly the orbital anisotropy of galaxies \citep[see][]{Binney05}.
This is achieved in Paper~X, in which a trend between the anisotropy parameter
in the meridional plane and the average intrinsic ellipticity is revealed (see their Fig.~7),
intrinsically more {\em flattened} galaxies tending to be more {\em anisotropic}.
This contradicts the view that discy galaxies are nearly isotropic, and we should
therefore reexamine the relation between the isophotal shape as measured by $a_4 / a$
and the kinematic status of early-type galaxies.

KB96 presented a fairly significant correlation between the mean ellipticity
$\epsilon$ and $a_4/a$ for their sample of early-type
galaxies (their Fig.~3). They mentioned that galaxies in a $\epsilon$
versus $a_4/a$ diagram exhibit a V-shaped distribution. The authors
therefore pointed out that this remarkable correlation implies that early-type
galaxies with nearly elliptical isophotes are also nearly round, and galaxies
with the most extreme $a_4 / a$ values are also the most flattened
\citep[see also][]{Hao06+}. We confirm this result using our photometric data on
the \Sauron\ sample as shown in Fig.~\ref{fig:eps_a4} (upper panel).  As
mentioned above, all slow rotators with the exception of NGC\,4550 have
ellipticities $\epsilon_e$ lower than 0.3, and our expectation that
these galaxies should have isophotes close to pure ellipses is verified: 10 out
of 11 have $\left|(a_4/a)_e\right| < 1 \%$ (with 9 out of 11 having
$\left|(a_4/a)_e\right| < 0.5 \%$), the exception being
NGC\,3414, for which the relatively large positive $(a_4 / a)_e$ value 
comes from the remarkable polar-ring structure \citep{vanDriel00+}. 

In fact, most fast rotators have a {\em maximum} absolute $\left|(a_4/a)\right|$ value within 1~$R_e$ 
larger than 2\%, whereas most slow rotators (with the exceptions again of NGC\,3414 and NGC\,4550) 
have {\em maximum} absolute $\left|(a_4/a)\right|$ values less than 2\%.
This means that most fast rotators in the \Sauron\ sample of E and S0 galaxies exhibit
significantly non-elliptical isophotes, but most slow rotators do not.
There is a tendency for the fast rotators to have positive $a_4 / a$ (discy isophotes),
and most slow rotators (7 out of 12) have negative $a_4 / a$ (boxy isophotes)
We do not detect any global and significant correlation between the boxiness and the angular
momentum per unit mass as quantified by $\lambda_{R_e}$ in the galaxies of our sample, 
so that $\left|(a_4/a)\right|$ is clearly not a good proxy for rotation or angular momentum.
This seems partly in contradiction with the claim by KB96 that rotation is dynamically less
important in boxy than in discy early-type galaxies. The latter result probably originated
from the combination of two observed facts: firstly boxy galaxies are on average less flattened
(extremely flattened galaxies are very discy), and secondly the use of 
$(V/\sigma)^{\star}$ as a proxy for the amount of organised rotation in the stellar component,
which tends to underestimate the rotational support for less flattened galaxies.

\subsection{Luminosity and mass}
\label{sec:sersic}
In the left panel of Fig.~\ref{fig:LRMb}, we show the distribution of $\lambda_{R_e}$ 
as a function of absolute magnitude $M_B$ for the 48 \Sauron\ E/S0 galaxies. 
The three slowest rotators (NGC\,4486, NGC\,4374, NGC\,5846) are among the brightest galaxies 
in our sample with $M_B < -21$~mag. Other slow rotators tend to be bright
but are spread over a wide range of absolute magnitude going from the relatively 
faint NGC\,4458, to brighter objects such as NGC\,5813. The bright and
faint end of slow rotators can be distinguished by the shapes of their
isophotes: slow rotators brighter than $M_B < -20$~mag all exhibit mildly boxy isophotes 
(negative $a_4 / a$ with amplitude less than 1\%) while the four discy slow 
rotators are all fainter than $M_B > -20$~mag, following the
known correlation between the isophote shapes and the total luminosity
of early-type galaxies \citep{BBF92}. Most fast rotators are fainter than $M_B > -20.5$~mag.

Going from total luminosity to mass, we have estimated the latter by approximating it with
the virial mass M$_{vir}$ derived from the best-fitting relation obtained in Paper~IV, namely
M$_{vir} \sim 5.0 \, R_e \sigma_e^2 / G$, where $\sigma_e$ is the luminosity
weighted second velocity moment within 1~$R_e$ (see Paper~IV for details).
The calculation of M$_{vir}$ from observables depends on the distance of the object,
which we obtained from different sources for the galaxies in our sample \citep[in
order of priority, from][and from the LEDA database assuming a Hubble flow with $H =
75$~km\,s$^{-1}$\,Mpc$^{-1}$]{Tonry+01,Tully88}.
A trend of $\lambda_{R_e}$ tending to be lower for more massive galaxies clearly emerges
if we now use this estimate of the virial mass M$_{vir}$ (right panel of Fig.~\ref{fig:LRMb}), 
as expected from the one observed with absolute magnitude $M_B$ (left panel of Fig.~\ref{fig:LRMb}).
The three slowest rotators are in the high range of M$_{vir}$ with values above $10^{11.5}$~M$_{\odot}$. 
There is a clear overlap in mass between fast and slow rotators for M$_{vir}$ between $10^{11}$
and $10^{11.5}$~M$_{\odot}$. However, all slow rotators, besides NGC\,4458 and NGC\,4550, 
have M$_{vir} > 10^{11}$~M$_{\odot}$, whereas most fast rotators have
M$_{vir} < 10^{11}$~M$_{\odot}$, lower masses being reached as the 
value of $\lambda_{R_e}$ increases. It is worth pointing out that the absolute
magnitude in the $K$ band, $M_K$, very nicely correlates with M$_{vir}$ (significantly better than
with $M_B$), so that the trend observed between $\lambda_R$ and M$_{vir}$ is also present if we were to
examine the relation between $\lambda_R$ and $M_K$.

Out of the 48 \Sauron\ galaxies, 33 have published cusp slope classification, distinguishing
``core" and ``power-law" galaxies \citep{Faber+97, Rest+01, Ravindranath+01, Lauer+05}. 
As illustrated in the left panel of Fig.~\ref{fig:LRMb}, we find that most slow
rotators are core galaxies, and most fast rotators are power-law galaxies
\citep[note that NGC\,821 has an $\lambda_{R_e} \sim 0.25$ and is specified as
an ``intermediate" object between a core and power-law in][]{Lauer+05}. 
There are yet no core galaxies with $\lambda_{R_e} > 0.3$, although the inclined galaxy NGC\,524
(see Paper~IV) would have a very high $\lambda_{R_e}$ value if seen edge-on.
These results are expected as there is 
a known trend between the central luminosity gradient and the total
luminosity of early-type galaxies, bright members tending to be core galaxies,
and lower luminosity ones to have power-law profiles \citep{Faber+97}.
All galaxies with $\lambda_{R_e} > 0.3$ have indeed $M_B > -20.7$ and 
conversely none have M$_{vir} > 10^{11.5}$~M$_{\odot}$.
This is, however, not a one-to-one correspondence since we find both
power-law in slow rotators (NGC\,3414 and NGC\,5813), and
core-like fast rotator (e.g., NGC\,524). There is also a domain in mass
and luminosity where both cusp and cores are found, namely for $M_{vir}$ between
$10^{11}$ and $10^{11.5}$~M$_{\odot}$, or correspondingly for $M_B$
from $-19.5$ to $-20.5$.
Another interesting result is found when examining the larger-scale luminosity profiles of
galaxies in our sample via the representation by a Sersic law. 
Besides the atypical case of NGC\,4550, all slow rotators have Sersic 
index $n > 4$: again, this is expected since galaxies
with larger Sersic shape index tend to be brighter \citep{Caon+93,GrahamGuzman03}.
Finally, galaxies with the lowest Sersic $n$ values are also 
among the fastest rotators. A more detailed account regarding these issues will
be provided in Falc{\'o}n-Barroso et al. (in preparation).

\subsection{Kinemetry groups}
\label{sec:kinresults}
\begin{figure}
\centering
\epsfig{file=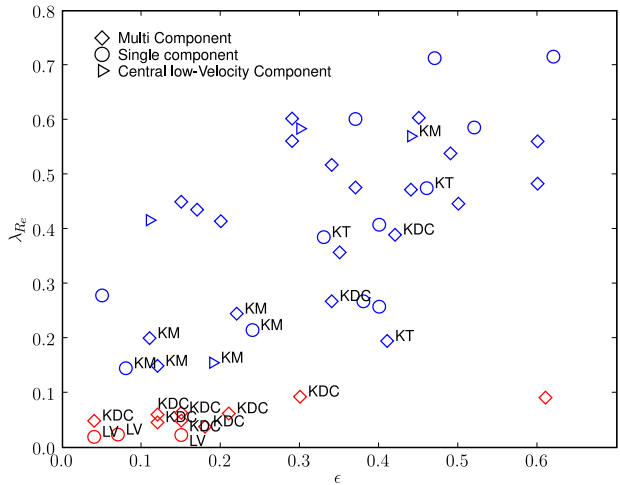, width=\columnwidth}
\caption{$\lambda_{R_e}$ versus mean ellipticity $\epsilon$, with indications
of the velocity structures identified in Sect.~\ref{sec:kinemetry}. Galaxies
with MCs are shown as diamonds, SCs as circles, and CLVs as right triangles. 
Colours for slow and fast rotators are as in Fig.~\ref{fig:RVS}.
KDC stand for kinematically decoupled components, KM and KT for kinematic
misalignment and twist, respectively, LV for Low Velocity component.}
\label{fig:KDC}
\end{figure}

We now turn to the kinemetric profiles (see Sect.~\ref{sec:kinemetry}) 
of the 48 \Sauron\ galaxies, which allow us to determine the number of observed kinematic 
components and their individual characteristics. We thus make use of the average photometric 
and kinematic position angles and axis ratio (PA$_{phot}$ and PA$_{kin}$, $q$ and $q_{kin}$), 
and the average and maximum of the velocity amplitude $k_1$.

The kinemetric groups of the observed 48 \Sauron\ velocity maps, defined in Sect.~\ref{sec:kinemetry}, are
provided in Table~\ref{tab:param}. Out of 48 galaxies in this sample, 33 (69\%) exhibit
multiple components (MC), including 10 kinematically decoupled components (KDCs; 21\%) and 4 
objects with central low-level velocity (CLVs; namely NGC\,3032, NGC\,4150, NGC\,4382, and NGC\,7457).
Two more galaxies with KDCs are in fact at the limit of being CLVs (NGC\,4621 and NGC\,7332) with
a maximum velocity for their inner component around 20~\kms\ at the \Sauron\ resolution. 
The latter two galaxies as well as the CLVs have in fact all been shown to harbour small
counter-rotating stellar systems \citep[][hereafter Paper~VIII]{WEC02,Falcon+04,PaperVIII}: it is the lower 
spatial resolution of the \Sauron\ data which produces the low central velocity gradient.
If we therefore count CLVs as galaxies with KDCs, this would result in a total of 14 KDC (29\%)
early-type objects.
Among the KDCs, 5 galaxies (NGC\,3414, NGC\,3608, NGC\,4458, NGC\,5813, NGC\,5831) have outer LV components.  
Kinematic misalignments are observed in 12 galaxies (25\%), and 6 galaxies have individual kinematic
components which have kinematical twists (KTs, 2 of them also having kinematical misalignments - KM).
Considering the difficulty of detecting these structures at certain viewing angles or low spatial resolution,
the number of such detected velocity structures is a lower limit.

In Fig.~\ref{fig:KDC} we show the $(\lambda_{R_e}, \epsilon)$ diagram (as in 
Fig.~\ref{fig:RVS}), now with the characteristic velocity structures
detected via kinemetry. The three slowest rotators 
are obviously tagged as low-velocity (LV) systems.
As mentioned above, these three galaxies are giant (bright and massive) 
roundish ellipticals with $M_B < -21$. Apart from these three and the atypical case of NGC\,4550, 
all other slow rotators harbour KDCs, the central kinematic
component having a typical size of 1~kpc or larger (see Paper~VIII). 
This contrasts with KDCs and CLVs in fast rotators (NGC\,3032, NGC\,4150, NGC\,7332, NGC\,7457),
where the size of the central (counter-rotating) stellar component is only a few arcseconds in radius, 
corresponding to significantly less than $500$~pc (Paper~VIII). 

A number of fast rotators show kinematic misalignments (KM) and twists (KT), most
of these being relatively small in contrast with the ones found in slow
rotators (see Sect.~\ref{sec:misalign}).
The four fastest rotators with a KT, KM,
or KDC, namely NGC\,1023, NGC\,3377, NGC\,7332 and NGC\,7457, are all barred galaxies.
Most of the galaxies with multi-components but no specific velocity structures
are among the fastest rotators. This result may be partly understood if these objects tend
to be close to edge-on, which renders the detection of such velocity structures harder,
but the detection of multiple components easier. 

\section{Discussion}
\label{sec:discussion}

The fast rotators are mostly discy galaxies exhibiting multiple components in
their stellar velocity fields. Contrarily to slow rotators, the main stellar
kinematic axis in fast rotators is relatively well aligned with the photometric 
major-axis, except for one galaxy (NGC\,474) known to harbour irregular shells. This result together
with the fact that the $\lambda_R$ profiles are qualitatively different for slow and
fast rotators clearly show that slow rotators are not velocity scaled-down versions of 
fast rotators. The small number of galaxies in our
sample, as well as our biased representation of the galaxy luminosity function,
reminds us that a larger and complete sample is required to reveal the true
$\lambda_{R_e}$ distribution in early-type galaxies. 
\begin{figure}
\centering
\epsfig{file=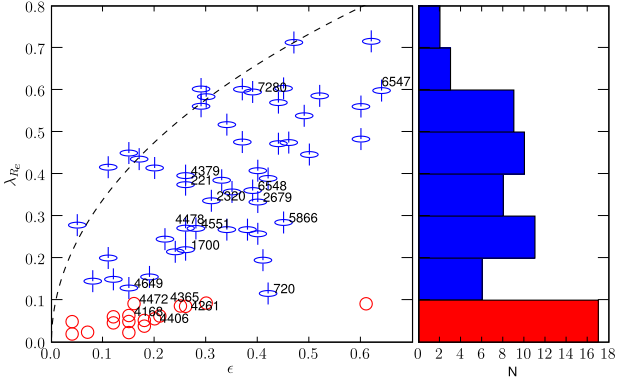, width=\columnwidth}
\caption{Left panel: $\lambda_{R_e}$ versus the ellipticity $\epsilon$ 
including the additional 18 E and S0 galaxies observed with \Sauron\ (labelled with NGC numbers). 
Symbols and the dashed line are as in Fig.~\ref{fig:RVS}. Right panel: histogram of $\lambda_{R_e}$
values including the 18 ``specials".}
\label{fig:RVSspec}
\end{figure}

\begin{figure}
\centering
\epsfig{file=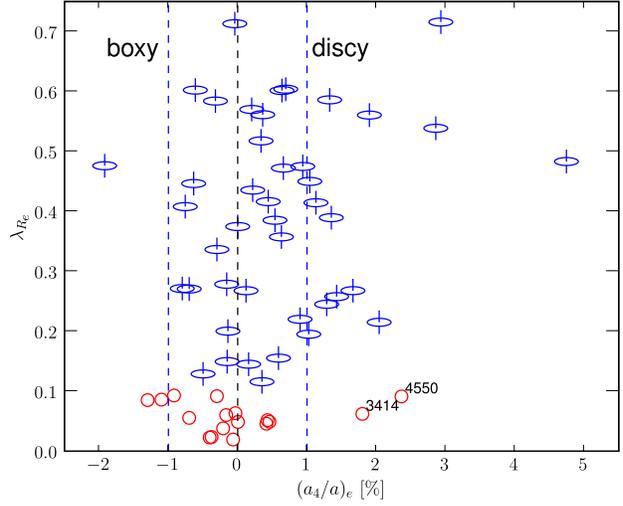, width=\columnwidth}
\caption{$\lambda_{R_e}$ versus $a_4$ for the 48 E/S0 of the \Sauron\ sample and an additional
12 galaxies for which we have available $a_4$ values. 
Symbols for slow and fast rotators are as in Fig.~\ref{fig:eps_a4}.}
\label{fig:LR_a4_18}
\end{figure}

\begin{figure}
\centering
\epsfig{file=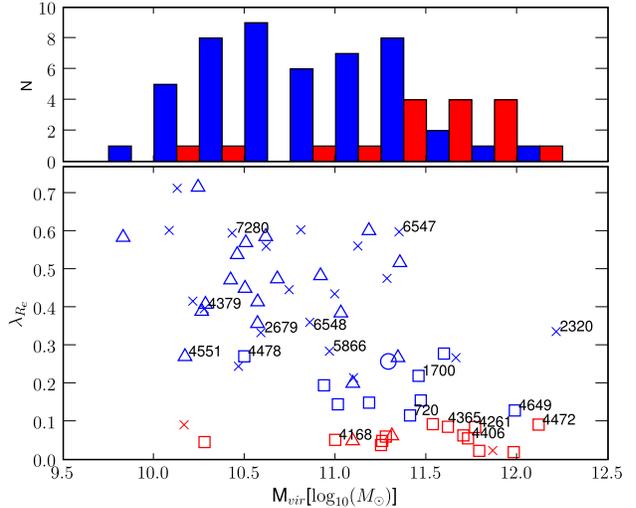, width=\columnwidth}
\caption{$\lambda_{R_e}$ versus the virial mass M$_{vir}$ 
for the 48 E/S0 of the \Sauron\ sample plus an additional
17 "specials" (we do not include the fast rotator NGC\,221 - M\,32 - 
in this plot considering its very low mass). The top panel shows histograms of 
M$_{vir}$ (in log, with steps of 0.5) for both slow (red) and fast rotators (blue). 
Colours in both panels for slow and fast rotators are 
as in Fig.~\ref{fig:eps_a4}. NGC numbers are indicated
for all specials. Symbols for galaxies with cores and cusps are 
as in the left panel of Fig.~\ref{fig:LRMb}.}
\label{fig:LR_Mass_18}
\end{figure}

We therefore first briefly discuss here the results obtained so far, this time
including 18 additional early-type galaxies (``specials'') observed with \Sauron\ within the course of
other specific projects. For these additional targets, we had to retrieve the main
photometric parameters (e.g., $R_e$, $\epsilon$, $a_4 / a$) from the literature \citep[e.g.][]{BDM88,Faber+97}, 
the kinematic measurements ($\sigma_e$, $\lambda_{R_e}$) being derived from the available \Sauron\ data 
as for the 48 E/S0 galaxies of the main sample (only 12 galaxies out of the 18 specials have available
$a_4$ measurements). Adding these 18 galaxies does not change 
the overall distribution of fast and slow rotators in a $\lambda_{R_e}$ versus $\epsilon$ diagram,
as shown in Fig.~\ref{fig:RVSspec}. The fraction of slow rotators (17 out of 66) is still $\sim 25$\%. We also
confirm that most slow rotators have relatively small
ellipticities ($\epsilon < 0.3$), in contrast with fast rotators which span a wide range of flattening.
The most remarkable fact lies in the 
confirmation that within these 18 extra targets, there are 5 slow rotators (NGC\,4168, NGC\,4406, NGC\,4472, 
NGC\,4261, NGC\,4365), and they {\em all} contain large (kpc) scale KDCs. 
Similarly, in Fig.~\ref{fig:LR_a4_18} we confirm the lack of a global significant correlation 
between $\lambda_{R_e}$ and $a_4 / a$ as shown in Fig.~\ref{fig:eps_a4}. Again, about two thirds of the 
slow rotators are boxy (5 out of 17 are discy, and 1 has $a_4 /a = 0$ ), and very discy galaxies 
tend to be fast rotators with high $\lambda_{R_e}$ values. However, $a_4 / a$ is clearly not a good proxy 
for angular momentum. We then show in Fig.~\ref{fig:LR_Mass_18} the relation between $\lambda_{R_e}$ and M$_{vir}$ with
an additional 17 ``specials'' (excluding the compact galaxy NGC\,221 [M\,32] which would 
stretch the plot unnecessarily). Most slow rotators are massive galaxies with M$_{vir} > 10^{11}$~M$_{\odot}$, 
with still only two exceptions (NGC\,4550 and NGC\,4458), reinforcing the results from Fig.~\ref{fig:LR_Mass_18}.
The trend for more massive galaxies to have lower $\lambda_{R_e}$ values is still observed, with
an overlap of fast and slow rotators in the range $10^{11} - 10^{12}$~M$_{\odot}$.
The fast rotator NGC\,2320 seems to have a rather large $\lambda_{R_e}$ for its virial mass
(being in fact the most massive galaxy out of the 66). NGC\,2320 is the most distant galaxy of
this set (with $D \sim 83$~Mpc), and has a rather unusual molecular gas content for
an early-type galaxy: it exhibits an asymmetric molecular gas disc with a mass of
about $4\times10^9$~M$_{\odot}$, interpreted as a sign of recent accretion or 
dynamical perturbation \citep{Young+05}. 

In Fig.~\ref{fig:LR_Mass_18}, we also show the ``cusp / core'' classification as in Sect.~\ref{sec:sersic}.
Most fast rotators are still galaxies with cusps, and most slow rotators
are core galaxies. Conversely, massive galaxies tend to have cores, and smaller ones tend to 
exhibit cuspy central luminosity profiles: in fact all galaxies with M$_{vir} > 10^{11.5}$~M$_{\odot}$
have cores. There is a transition region, in the range $10^{11} - 10^{11.5}$~M$_{\odot}$ 
where we find both cusp and core galaxies. More interestingly,
all core galaxies still have $\lambda_{R_e} < 0.3$. However, a galaxy like NGC\,524 is almost 
certainly very inclined (see Paper~IV), its corresponding ``edge-on'' $\lambda_{R_e}$ 
value being then much larger than 0.75. It would thus be interesting to 
understand if an edge-on version of NGC\,524 would still have a core-like central
luminosity profile. We also observe a few galaxies with cusps and rather
low $\lambda_{R_e}$ values (NGC\,5831) which cannot be reconciled with inclined fast rotators.
We may therefore only discriminate cusp and core galaxies combining
both their mass {\em and} their $\lambda_{R_e}$ values: massive galaxies 
(M$_{vir} > 10^{11.5}$~M$_{\odot}$) with $\lambda_{R_e} < 0.3$ are indeed all core galaxies, 
while smaller galaxies (M$_{vir} < 10^{11}$~M$_{\odot}$) with $\lambda_{R_e} > 0.3$
seem to all be galaxies with cusps. The fact that we observe very few massive 
(M$_{vir} > 10^{11}$~M$_{\odot}$) galaxies with high $\lambda_{R_e}$ values implies
that either our sample is biased against them (e.g. we only observed close to face-on massive galaxies), 
or these galaxies are rare. Since it is difficult to understand how such a bias could affect our 
sample, we need to conclude that indeed massive galaxies tend to have little or no baryonic angular momentum
within one effective radius, and that the trend observed in Fig.~\ref{fig:LR_Mass_18} is real (see e.g., the
upper panel with the histogram of M$_{vir}$).

Interpreting the measured kinematic parameter $\lambda_R$ as a proxy for the
amount of stellar angular momentum per unit mass in the central region of early-type galaxies 
(see Appendix~A), we can then discuss the origin of this angular momentum.
The standard scenario for the formation of galaxy structures
includes hierarchical clustering of cold dark matter halos within which gas
is cooling \citep{Peebles69,Dor70,White84}. The angular momentum of the dark
matter halos is thought to originate in cosmological torques and
major mergers \citep[e.g.][]{Vivitska02+}, this
hypothesis providing a reasonable frame for the formation of disc galaxies.
If major mergers produce a significant increase in the 
specific angular momentum of the dark matter halos at large radii \citep{Vivitska02+}, 
minor mergers seem to just preserve or only slightly increase it with time
\citep{Donghia02}. But little is known on the expected distribution of the angular 
momentum of the {\em baryons} \citep{vdB02+,deJong04}, and even less if 
we focus on the central regions of galaxies (within a few effective radii).
In fact, discs formed in numerical simulations are generally an order of
magnitude too small \citep[but see][for a possible solution]{Dutton+06}.

Numerical simulations, whether they include a dissipative component or not, 
have helped us to understand how mergers influence
the rotational support of the baryonic component in galaxies
\citep[see e.g.][]{Barnes98,Somerville+99,Cole+00,Bournaud05+,Naab06+}. 
Stellar discs are cold and fragile
systems, thus easily destroyed during major mergers which often lead
to elliptical-like remnants even in the presence of a moderate amount of gas \citep{NJB06}. 
Intermediate to minor mergers preserve part of the disc better, but in most cases 
a merger leads to a redistribution of the angular momentum of the central
stellar component outwards \citep{Bournaud04+}. 

In this context, fast rotators have either preserved or regained their
specific angular momentum in the central part.
Since both very gas-rich major and minor mergers seem to produce fast (disc-dominated) 
rotators \citep{Robertson+06b, Cox+06}, this requires either the absence of a
major dry merger which would expel most of the angular momentum outwards,
or the rebuilding of a disc-like system via gas accretion.
At high redshift ($z > 2$), gas was abundant, and very gas-rich mergers should have
therefore been common. \cite{Robertson+06a} in fact claim that progenitors of
early-type galaxies must be gas-rich (gas fraction $> 30$\%) to produce the tilt of the
fundamental plane (FP). Dry major mergers are therefore expected to occur
preferentially at lower $z$, with fast rotators not having suffered from such events. 
In the picture of the hierarchical formation of structures, minor mergers 
are more common than major mergers and we can thus expect galaxies (slow and fast rotators)
to have suffered from more than one of these events up to $z=0$. 
\begin{figure}
\centering
\epsfig{file=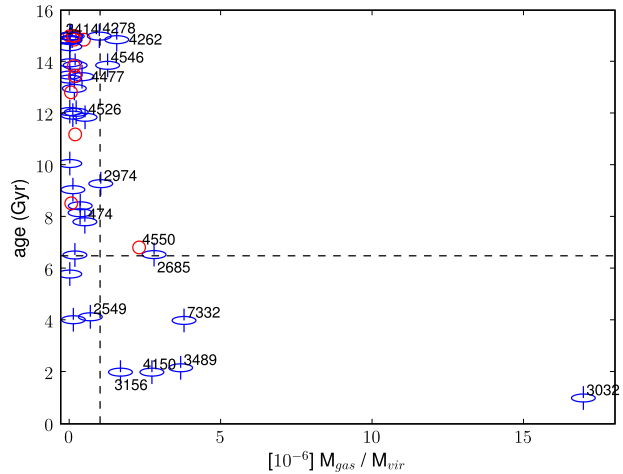, width=\columnwidth}
\caption{Average luminosity-weighted age of the stellar component versus the
specific mass of ionised gas within the \Sauron\ field of view M$_{gas}$/M$_{vir}$
(ionised gas mass normalised by 
the virial mass, see Sect.~\ref{sec:sersic}). Symbols for fast and slow rotators 
are as in Fig.~\ref{fig:RVS}. Numbers for the ionised gas specific mass
are derived from values in Papers~IV and Paper~V, and average luminosity-weighted 
ages are from Kuntschner et al. (in preparation). The vertical and horizontal dashed
lines shows the limits of $> 10^{-6}$~M$_{gas}$/M$_{vir}$ and 6.5~Gyr ($z\sim 1$),
respectively.}
\label{fig:LRage}
\end{figure}

The resulting specific angular momentum of fast rotators, as quantified by
$\lambda_R$, is therefore expected to result mostly 
from a competition between (i) gas-rich minor mergers or other
inflow of external gas that causes a gradual increase of $\lambda_R$ (in the
inner parts), and (ii) dissipationless dry minor mergers triggering
disc instability and heating, and resulting in the transformation of disc
material into a more spheroidal component, lowering $\lambda_R$ (in the inner parts).  
The scenario described here follows the idea previously sketched by many authors
\citep[e.g.][]{BBF92, KB96, Faber+97,NJB06} that dissipation is important in the
formation process of fast rotating early-type galaxies. The `heating' of 
the disc via star formation leads to an increase in the vertical dispersion, more
isotropic and rounder galaxies, and therefore moving the galaxies
along the anisotropy--ellipticity trend (Paper~X) towards the slow
rotators. A sudden removal of the gas (e.g., due to AGN feedback or
ram pressure stripping) might have `quenched' this process, and
quickly moved the galaxy from the `blue cloud' to the `red sequence' \citep{Faber06}.
Some galaxies may still have recently accreted some gas \citep[e.g. via
interaction with a companion, see][]{Falcon+04}, and sometimes show the presence of a younger
stellar population \citep[such as NGC\,3032, NGC\,3489, NGC\,4150, see][Paper~VI, and
see also Paper~VIII]{PaperVI}.
In fact all \Sauron\ E/S0 galaxies which have luminosity weighted ages of their stellar
component lower than $6.5$~Gyr ($z\sim1$) are fast rotators (Fig.~\ref{fig:LRage}; 
see also Kuntschner et al. in preparation).
Among the 11 galaxies in our sample with a high ionised gas content
($\mbox{M}_{gas}/\mbox{M}_{vir}> 10^{-6}$), only one is a slow rotator, namely 
the atypical disc galaxy NGC\,4550. The slow rotator with the second highest gas content
is the polar-ring galaxy NGC\,3414 (M$_{gas}$/M$_{vir} \sim 0.6 \times 10^{-6}$), 
NGC\,4550 and NGC\,3414 being in fact the two galaxies having 
deviant $(a_4/a)_e$ values with respect to all other slow rotators 
(see Sect.~\ref{sec:isophotes}).

Following the same line of argument, slow rotators need to have expelled most of their 
angular momentum within one effective radius outwards. This would require a
significant fraction of the mass to be accreted during mergers with relatively
gas-poor content. As emphasised above, all slow rotators with $\lambda_{R_e} > 0.03$
(thus excluding the three slowest rotators which are consistent with zero rotation) 
have stellar kpc-size KDCs. Opposite to the smaller-scale KDCs in fast rotators
(Sect.~\ref{sec:kinresults}), the latter have a similar old stellar population as the
rest of the galaxy (Paper~VIII), and thus were formed long ago when
gas was more abundant than today. Indeed, numerical simulations of the formation of
such structures seem to require the presence of some gas during
the merger process followed by subsequent star formation, 
either in equal-mass mergers \citep{NJB06} or
in multiple intermediate or minor mergers (Bournaud, PhD Thesis, Paris).
Fig.~\ref{fig:LRage} reveals that most slow rotators
have a relatively modest fraction of ionised gas and 
are in fact older than $\sim 6.5$~Gyr ($z\sim 1$).
The only two exceptions are, again, NGC\,4550 which is very probably the
result of the rare encounter of two disc galaxies 
with nearly exactly opposite spins (see Paper~X), and 
NGC\,3414 which shows the ``rudiments of a disc" \citep{Carnegie94}, and
photometric structures reminiscent of polar-ring galaxies \citep{vanDriel00+}
presumably formed via a tidal gas accretion event \citep{Bournaud+03}.
These results are consistent with the suggestion that slow rotators did not suffer 
from significant (in mass) recent dry mergers which may destroy or diffuse the large-scale KDCs. 
Minor mergers may still have recently occurred, and as above result in a competition of (slightly)
lowering and increasing $\lambda_R$.

The three slowest rotators (NGC\,4374, NGC\,4486, and NGC\,5846) may then correspond to the extreme end point 
of early-type galaxy evolution where dry mergers had a major role in the evolution,
allowing $\lambda_R$ to reach values consistent with zero in the central region. These galaxies
are found in rather dense galactic environment and are expected to have experienced a significant
number of mergers with galaxies which were
themselves gas-depleted (due to ram-pressure stripping, due to efficient 
AGN feedback, or due to earlier mergers),
preferentially on radial orbits along the cosmological filaments \citep{Boylan+06}.
As for the other slow-rotators, the initial mergers will have been gas-rich to
place them on the fundamental plane, but later-on dry mergers are expected to be frequent as
galaxies fall into the cluster potential well. The insignificant role of star
formation in the late history of these galaxies is reflected in both their
low specific gas content and old luminosity weighted ages (Fig.~\ref{fig:LRage}).
Overall, the trend we observe for early-type galaxies to have lower $\lambda_R$ 
with increasing mass (Fig.~\ref{fig:LRMb})
is consistent with the gradually more important role of (gas-poor) mergers when going
from fast rotators, to slow rotators. 

Is this also consistent with the observed
trend between $\lambda_{R_e}$, M$_{vir}$ and the cusp slope described above?
Cores are thought to be the result of the scouring via a binary supermassive black hole
\citep[][and references therein]{Faber+97, Merritt06}, the potential consequence of a
merger. After a core is formed, a subsequent central accretion of gaseous material 
(from external or internal sources) followed by an episode of star formation
could rejuvenate a cuspy luminosity profile, most probably in the form of an inner disc. 
If this scenario is correct, there should be a relation between 
the presence of a core and $\lambda_{R_e}$. The observation that massive slow rotators, thought
to have significantly suffered from dissipationless mergers, 
all exhibit cores, and that small galaxies with high $\lambda_{R_e}$ all 
contain cusps does indeed fit into this scenario. However, the fact that
minor gas-free mergers may only remove part of the baryonic angular momentum,
and that gas can be accreted from e.g., external sources or stellar mass loss, 
implies that there should not be a one-to-one relation, as indeed
observed. Intriguingly, the two slow rotators with cusps (NGC\,3414, NGC\,5813) both 
show the presence of a decoupled stellar disc-like component in their central few hundreds of parsecs
(as witnessed by the amplitude of the associated $h_3$ parameter, see Paper~III).
These may illustrate again the competition between
gas-poor mergers and gas accreting processes in the making of galaxy structures.

\section{Conclusions}
\label{sec:conclusions}
Using two-dimensional stellar kinematics, we have argued that
early-type galaxies can be divided into two broad dynamical classes. 
We have devised a new parameter $\lambda_R$ to quantify the specific
angular momentum in the stellar component of galaxies. This parameter,
as applied within one effective radius to the $48$ E and S0 galaxies
of our \Sauron\ sample, allowed us to disentangle the two classes
of fast and slow rotators with a threshold value of $0.1$. 
As emphasised in Sect.~\ref{sec:rotators}, fast and slow rotators exhibit
qualitatively and quantitatively different stellar kinematics (see also Paper~X). 
Galaxies with $\lambda_R \le 0.1$ form a
distinct class characterised by little or no large-scale rotation, by the
presence of kinematically decoupled cores, and by significant 
kinematic misalignments and/or velocity twists.
This contrasts with galaxies for which $\lambda_R > 0.1$, which display global rotation 
along a well defined apparent kinematic major-axis that is nearly aligned with the photometric 
major-axis. We have also shown that slow and fast rotators exhibit qualitatively different $\lambda_R$ 
radial profiles and that slow-rotators are unlikely to be face-on versions of fast-rotators.

Beside the obvious difference in their angular momentum content, the
main properties of the fast and slow rotators are as follows:
\renewcommand{\labelenumi}{(\roman{enumi})}
\begin{enumerate}
\item 75\% of the galaxies in our sample are fast rotators, 25\% are slow
rotators. Most of the slow rotators are classified as Es, while
fast rotators include a mix of Es and S0s.
\item Within the slow rotators, three early-type galaxies have velocity fields
consistent with near zero rotation. These three objects (NGC\,4374$\equiv$M\,84,
NGC\,4486$\equiv$M\,87 and NGC\,5846) are bright and massive, nearly round galaxies. 
\item Fast rotators have ellipticities up to $\epsilon \sim 0.6$, 
in contrast to slow rotators which are relatively round systems with $\epsilon < 0.3$.
\item Fast rotators tend to be relatively low-luminosity objects with $M_B > -20.7$.
Slow rotators cover almost the full range of absolute magnitude of our
sample, from faint discy galaxies such as NGC\,4458 to bright slightly boxy
galaxies such as NGC\,5813, although they on average tend to be brighter than fast rotators.
All slow rotators have Sersic index $n$ larger than 4, except the atypical galaxy NGC\,4550.
\item More than 70\% of the galaxies in our sample exhibit multi-component
kinematic systems. As mentioned above, fast rotators have nearly aligned photometric and kinematic
components, except for NGC\,474 which exhibits obvious irregular shells. 
This contrasts with slow rotators which have significant
kinematic misalignments and/or velocity twists.
\item All slow rotators, besides the three slowest rotators (for which stellar velocities are 
consistent with zero everywhere), 
exhibit a kiloparcsec-scale stellar KDC (see also Paper~VIII). 
\end{enumerate}

These results are confirmed when including 18 additional galaxies observed
with \Sauron. Although our sample of 48 early-type galaxies is not a good representation 
of the galaxy luminosity distribution in the nearby Universe, it is the first to provide such
a constraint for numerical simulations of galaxy formation and evolution.
It is clear that a significantly larger and unbiased sample is
required to probe the true distribution of $\lambda_R$ in early-type galaxies
and the fraction of slow and fast rotators.
Even so, our sample should serve as a reference point for detailed numerical
simulations of galactic systems in a cosmological context.

\section*{Acknowledgements}

EE warmly thank Frank van den Bosch for insightful discussions.  The \Sauron\ project is made possible
through grants 614.13.003, 781.74.203, 614.000.301 and 614.031.015 from the 
Nederlandse Organisatie voor WettenSchappelijk OnderzoekNetherlands (NWO) and
financial contributions from the Institut National des Sciences de l'Univers,
the Universit\'e Lyon~I, the Universities of Durham,
Leiden, and Oxford, the Programme National Galaxies, the British Council, PPARC grant `Observational Astrophysics 
at Oxford 2002--2006' and support from Christ Church Oxford, and the Netherlands
Research School for Astronomy NOVA.  RLD is grateful for the award of
a PPARC Senior Fellowship (PPA/Y/S/1999/00854) and postdoctoral
support through PPARC grant PPA/G/S/2000/00729. The PPARC Visitors
grant (PPA/V/S/2002/00553) to Oxford also supported this work. 
MC acknowledges support from a VENI award (639.041.203) by NWO and a 
PPARC Advanced Fellowship (PP/D005574/1)
GvdV acknowledges support provided by NASA through grant NNG04GL47G
and through Hubble Fellowship grant HST-HF-01202.01-A awarded by the
Space Telescope Science Institute, which is operated by the
Association of Universities for Research in Astronomy, Inc., for
NASA, under contract NAS 5-26555. JFB
acknowledges support from the Euro3D Research Training Network, funded
by the EC under contract HPRN-CT-2002-00305. This paper is based on observations obtained
at the William Herschel Telescope, operated by the 
Isaac Newton Group in the Spanish Observatorio del
Roque de los Muchachos of the Instituto de Astrof\'{\i}sica de
Canarias. This project made use of the HyperLeda (http://leda.univ-lyon1.fr), and NED databases. 
Part of this work is based on data
obtained from the ESO/ST-ECF Science Archive Facility. Photometric
data were obtained (in part) using the 1.3m McGraw-Hill Telescope of
the MDM Observatory. 
\bibliographystyle{mn2e}
\bibliography{Paper9_published_astroph}

\begin{thebibliography}{}

\bibitem[\protect\citeauthoryear{{Bacon}, {Adam}, {Baranne}, {Courtes},
  {Dubet}, {Dubois}, {Emsellem}, {Ferruit}, {Georgelin}, {Monnet}, {Pecontal},
  {Rousset} \& {Say}}{{Bacon} et~al.}{1995}]{Bacon+95}
{Bacon} R.,  {Adam} G.,  {Baranne} A.,  {Courtes} G.,  {Dubet} D.,  {Dubois}
  J.~P.,  {Emsellem} E.,  {Ferruit} P.,  {Georgelin} Y.,  {Monnet} G.,
  {Pecontal} E.,  {Rousset} A.,    {Say} F.,  1995, \aaps, 113, 347

\bibitem[\protect\citeauthoryear{{Bacon}, {Copin}, {Monnet}, {Miller},
  {Allington-Smith}, {Bureau}, {Carollo}, {Davies}, {Emsellem}, {Kuntschner},
  {Peletier}, {Verolme} \& {de Zeeuw}}{{Bacon} et~al.}{2001}]{Bacon+01}
{Bacon} R.,  {Copin} Y.,  {Monnet} G.,  {Miller} B.~W.,  {Allington-Smith}
  J.~R.,  {Bureau} M.,  {Carollo} C.~M.,  {Davies} R.~L.,  {Emsellem} E.,
  {Kuntschner} H.,  {Peletier} R.~F.,  {Verolme} E.~K.,    {de Zeeuw} P.~T.,
  2001, \mnras, 326, 23 (Paper~I)

\bibitem[\protect\citeauthoryear{{Barnes}}{{Barnes}}{1998}]{Barnes98}
{Barnes} J.~E.,  1998, Saas-Fee Advanced Course 26. Lecture Notes 1996. Swiss Society for Astrophysics and Astronomy, XIV, Edited by R. C. Kennicutt, Jr. F. Schweizer, J. E. Barnes, D. Friedli, L. Martinet, and D. Pfenniger. 404 pp., 214 figs. Springer-Verlag Berlin/Heidelberg; ISBN: 3-540-63569-6, 1998., p.275

\bibitem[\protect\citeauthoryear{{Bender}}{{Bender}}{1988}]{Bender88}
{Bender} R.,  1988, \aap, 202, L5

\bibitem[\protect\citeauthoryear{{Bender}, {Burstein} \& {Faber}}{{Bender}
  et~al.}{1992}]{BBF92}
{Bender} R.,  {Burstein} D.,    {Faber} S.~M.,  1992, \apj, 399, 462

\bibitem[\protect\citeauthoryear{{Bender}, {Doebereiner} \&
  {Moellenhoff}}{{Bender} et~al.}{1988}]{BDM88}
{Bender} R.,  {Doebereiner} S.,    {Moellenhoff} C.,  1988, \aaps, 74, 385

\bibitem[\protect\citeauthoryear{{Bender} \& {Moellenhoff}}{{Bender} \&
  {Moellenhoff}}{1987}]{BM87}
{Bender} R.,  {Moellenhoff} C.,  1987, \aap, 177, 71

\bibitem[\protect\citeauthoryear{{Bender}, {Saglia} \& {Gerhard}}{{Bender}
  et~al.}{1994}]{BSG94}
{Bender} R.,  {Saglia} R.~P.,    {Gerhard} O.~E.,  1994, \mnras, 269, 785

\bibitem[\protect\citeauthoryear{{Binney}}{{Binney}}{1978}]{Binney78}
{Binney} J.,  1978, \mnras, 183, 501

\bibitem[\protect\citeauthoryear{{Binney}}{{Binney}}{2005}]{Binney05}
{Binney} J.,  2005, \mnras, 363, 937

\bibitem[\protect\citeauthoryear{{Bournaud} \& {Combes}}{{Bournaud} \&
  {Combes}}{2003}]{Bournaud+03}
{Bournaud} F.,  {Combes} F.,  2003, \aap, 401, 817

\bibitem[\protect\citeauthoryear{{Bournaud}, {Combes} \& {Jog}}{{Bournaud}
  et~al.}{2004}]{Bournaud04+}
{Bournaud} F.,  {Combes} F.,    {Jog} C.~J.,  2004, \aap, 418, L27

\bibitem[\protect\citeauthoryear{{Bournaud}, {Jog} \& {Combes}}{{Bournaud}
  et~al.}{2005}]{Bournaud05+}
{Bournaud} F.,  {Jog} C.~J.,    {Combes} F.,  2005, \aap, 437, 69

\bibitem[\protect\citeauthoryear{{Boylan-Kolchin}, {Ma} \&
  {Quataert}}{{Boylan-Kolchin} et~al.}{2006}]{Boylan+06}
{Boylan-Kolchin} M.,  {Ma} C.-P.,    {Quataert} E.,  2006, \mnras, 369, 1081

\bibitem[\protect\citeauthoryear{{Burkert} \& {Naab}}{{Burkert} \&
  {Naab}}{2005}]{BN05}
{Burkert} A.,  {Naab} T.,  2005, \mnras, 363, 597

\bibitem[\protect\citeauthoryear{{Caon}, {Capaccioli} \& {D'Onofrio}}{{Caon}
  et~al.}{1993}]{Caon+93}
{Caon} N.,  {Capaccioli} M.,    {D'Onofrio} M.,  1993, \mnras, 265, 1013

\bibitem[\protect\citeauthoryear{{Cappellari}, {Bacon}, {Bureau}, {Damen},
  {Davies}, {de Zeeuw}, {Emsellem}, {Falc{\'o}n-Barroso}, {Krajnovi{\'c}},
  {Kuntschner}, {McDermid}, {Peletier}, {Sarzi}, {van den Bosch} \& {van de
  Ven}}{{Cappellari} et~al.}{2006}]{PaperIV}
{Cappellari} M.,  {Bacon} R.,  {Bureau} M.,  {Damen} M.~C.,  {Davies} R.~L.,
  {de Zeeuw} P.~T.,  {Emsellem} E.,  {Falc{\'o}n-Barroso} J.,  {Krajnovi{\'c}}
  D.,  {Kuntschner} H.,  {McDermid} R.~M.,  {Peletier} R.~F.,  {Sarzi} M.,
  {van den Bosch} R.~C.~E.,    {van de Ven} G.,  2006, \mnras, 366, 1126 (Paper~IV)

\bibitem[\protect\citeauthoryear{{Cappellari} \& {Copin}}{{Cappellari} \&
  {Copin}}{2003}]{CC03}
{Cappellari} M.,  {Copin} Y.,  2003, \mnras, 342, 345

\bibitem[\protect\citeauthoryear{{Cappellari} \& {Emsellem}}{{Cappellari} \&
  {Emsellem}}{2004}]{CE04}
{Cappellari} M.,  {Emsellem} E.,  2004, \pasp, 116, 138

\bibitem[\protect\citeauthoryear{{Cappellari}, {Emsellem}, {Bacon}, {Bureau},
  {Davies}, {de Zeeuw}, {Falc{\'o}n-Barroso}, {Krajnovi{\'c}}, {Kuntschner},
  {McDermid}, {Peletier}, {Sarzi}, {van den Bosch} \& {van de
  Ven}}{{Cappellari} et~al.}{2007}]{PaperX}
{Cappellari} M.,  {Emsellem} E.,  {Bacon} R.,  {Bureau} M.,  {Davies} R.~L.,
  {de Zeeuw} P.~T.,  {Falc{\'o}n-Barroso} J.,  {Krajnovi{\'c}} D.,
  {Kuntschner} H.,  {McDermid} R.~M.,  {Peletier} R.~F.,  {Sarzi} M.,  {van den
  Bosch} R. C.~E.,    {van de Ven} G.,  2007, \mnras, in press (Paper~X) (astro-ph/0703533)

\bibitem[\protect\citeauthoryear{{Cole}, {Lacey}, {Baugh} \& {Frenk}}{{Cole}
  et~al.}{2000}]{Cole+00}
{Cole} S.,  {Lacey} C.~G.,  {Baugh} C.~M.,    {Frenk} C.~S.,  2000, \mnras,
  319, 168

\bibitem[\protect\citeauthoryear{{Cox}, {Dutta}, {Di Matteo}, {Hernquist},
  {Hopkins}, {Robertson} \& {Springel}}{{Cox} et~al.}{2006}]{Cox+06} 
{Cox} T.~J.,  {Dutta} S.~N.,  {Di Matteo} T.,  {Hernquist} L.,  {Hopkins}
  P.~F.,  {Robertson} B.,    {Springel} V.,  2006, \apj, 650, 791 

\bibitem[\protect\citeauthoryear{{Davies}, {Efstathiou}, {Fall}, {Illingworth}
  \& {Schechter}}{{Davies} et~al.}{1983}]{Davies+83}
{Davies} R.~L.,  {Efstathiou} G.,  {Fall} S.~M.,  {Illingworth} G.,
  {Schechter} P.~L.,  1983, \apj, 266, 41

\bibitem[\protect\citeauthoryear{{de Jong}, {Kassin}, {Bell} \& {Courteau}}{{de
  Jong} et~al.}{2004}]{deJong04}
{de Jong} R.~S.,  {Kassin} S.,  {Bell} E.~F.,    {Courteau} S.,  2004, in
the proceedings of the International Astronomical Union Symposium no. 220, held 21 - 25 July, 
   2003 in Sydney, Australia. Eds: S. D. Ryder, D. J. Pisano, M. A. Walker, 
   and K. C. Freeman. San Francisco: Astronomical Society of the Pacific., p.281

\bibitem[\protect\citeauthoryear{{de Jong}, {Simard}, {Davies}, {Saglia},
  {Burstein}, {Colless}, {McMahan} \& {Wegner}}{{de Jong}
  et~al.}{2004}]{deJong+04}
{de Jong} R.~S.,  {Simard} L.,  {Davies} R.~L.,  {Saglia} R.~P.,  {Burstein}
  D.,  {Colless} M.,  {McMahan} R.,    {Wegner} G.,  2004, \mnras, 355, 1155

\bibitem[\protect\citeauthoryear{{de Vaucouleurs}, {de Vaucouleurs}, {Corwin}
  Jr., {Buta}, {Paturel} \& {Fouque}}{{de Vaucouleurs} et~al.}{1991}]{RC3}
{de Vaucouleurs} G.,  {de Vaucouleurs} A.,  {Corwin} Jr. H.~G.,  {Buta} R.~J.,
  {Paturel} G.,    {Fouque} P.,  1991, {Third Reference Catalogue of Bright
  Galaxies}.
Volume 1-3, XII, 2069 pp.~7 figs..~ Springer-Verlag Berlin Heidelberg New York

\bibitem[\protect\citeauthoryear{{de Zeeuw}, {Bureau}, {Emsellem}, {Bacon},
  {Carollo}, {Copin}, {Davies}, {Kuntschner}, {Miller}, {Monnet}, {Peletier} \&
  {Verolme}}{{de Zeeuw} et~al.}{2002}]{PaperII}
{de Zeeuw} P.~T.,  {Bureau} M.,  {Emsellem} E.,  {Bacon} R.,  {Carollo} C.~M.,
  {Copin} Y.,  {Davies} R.~L.,  {Kuntschner} H.,  {Miller} B.~W.,  {Monnet} G.,
   {Peletier} R.~F.,    {Verolme} E.~K.,  2002, \mnras, 329, 513 (Paper~II)

\bibitem[\protect\citeauthoryear{{D'Onghia} \& {Burkert}}{{D'Onghia} \&
  {Burkert}}{2004}]{Donghia02}
{D'Onghia} E.,  {Burkert} A.,  2004, \apjl, 612, L13

\bibitem[\protect\citeauthoryear{{Doroshkevich}}{{Doroshkevich}}{1970}]{Dor70}
{Doroshkevich} A.~G.,  1970, Astrophysics, 6, 320

\bibitem[\protect\citeauthoryear{{Dutton}, {van den Bosch}, {Dekel} \&
  {Courteau}}{{Dutton} et~al.}{2006}]{Dutton+06}
{Dutton} A.~A.,  {van den Bosch} F.~C.,  {Dekel} A.,    {Courteau} S.,  2007, \apj, 654, 27

\bibitem[\protect\citeauthoryear{{Emsellem}, {Cappellari}, {Peletier},
  {McDermid}, {Bacon}, {Bureau}, {Copin}, {Davies}, {Krajnovi{\'c}},
  {Kuntschner}, {Miller} \& {de Zeeuw}}{{Emsellem} et~al.}{2004}]{Emsellem+04}
{Emsellem} E.,  {Cappellari} M.,  {Peletier} R.~F.,  {McDermid} R.~M.,  {Bacon}
  R.,  {Bureau} M.,  {Copin} Y.,  {Davies} R.~L.,  {Krajnovi{\'c}} D.,
  {Kuntschner} H.,  {Miller} B.~W.,    {de Zeeuw} P.~T.,  2004, \mnras, 352,
  721

\bibitem[\protect\citeauthoryear{{Faber}, {Tremaine}, {Ajhar}, {Byun},
  {Dressler}, {Gebhardt}, {Grillmair}, {Kormendy}, {Lauer} \&
  {Richstone}}{{Faber} et~al.}{1997}]{Faber+97}
{Faber} S.~M.,  {Tremaine} S.,  {Ajhar} E.~A.,  {Byun} Y.-I.,  {Dressler} A.,
  {Gebhardt} K.,  {Grillmair} C.,  {Kormendy} J.,  {Lauer} T.~R.,
  {Richstone} D.,  1997, \aj, 114, 1771

\bibitem[\protect\citeauthoryear{{Faber}, {Willmer}, {Wolf}, {Koo}, {Weiner} \&
  {Newman} J.~A.}{{Faber} et~al.}{2005}]{Faber06}
{Faber} S.~M.,  {Willmer} C.~N.~A.,  {Wolf} C.,  {Koo} D.~C.,  {Weiner} B.~J.,
    {Newman} J.~A. e.~a.,  2005, ArXiv Astrophysics e-prints

\bibitem[\protect\citeauthoryear{{Falc{\'o}n-Barroso}, {Peletier}, {Emsellem},
  {Kuntschner}, {Fathi}, {Bureau}, {Bacon}, {Cappellari}, {Copin}, {Davies} \&
  {de Zeeuw}}{{Falc{\'o}n-Barroso} et~al.}{2004}]{Falcon+04}
{Falc{\'o}n-Barroso} J.,  {Peletier} R.~F.,  {Emsellem} E.,  {Kuntschner} H.,
  {Fathi} K.,  {Bureau} M.,  {Bacon} R.,  {Cappellari} M.,  {Copin} Y.,
  {Davies} R.~L.,    {de Zeeuw} P.~T.,  2004, \mnras, 350, 35

\bibitem[\protect\citeauthoryear{{Ferrarese}, {C{\^o}t{\'e}}, {Jord{\'a}n},
  {Peng}, {Blakeslee}, {Piatek}, {Mei}, {Merritt}, {Milosavljevi{\'c}}, {Tonry}
  \& {West}}{{Ferrarese} et~al.}{2006}]{ACSVI}
{Ferrarese} L.,  {C{\^o}t{\'e}} P.,  {Jord{\'a}n} A.,  {Peng} E.~W.,
  {Blakeslee} J.~P.,  {Piatek} S.,  {Mei} S.,  {Merritt} D.,
  {Milosavljevi{\'c}} M.,  {Tonry} J.~L.,    {West} M.~J.,  2006, \apjs, 164,
  334

\bibitem[\protect\citeauthoryear{{Fisher}}{{Fisher}}{1997}]{Fisher97}
{Fisher} D.,  1997, \aj, 113, 950

\bibitem[\protect\citeauthoryear{{Franx}}{{Franx}}{1988}]{Franx88}
{Franx} M.,  1988, {Structure and kinematics of elliptical galaxies}.
Leiden: Rijksuniversiteit, 1988

\bibitem[\protect\citeauthoryear{{Franx}, {Illingworth} \& {Heckman}}{{Franx}
  et~al.}{1989}]{Franx+89}
{Franx} M.,  {Illingworth} G.,    {Heckman} T.,  1989, \aj, 98, 538

\bibitem[\protect\citeauthoryear{{Fritze v.~Alvensleben}}{{Fritze
  v.~Alvensleben}}{2004}]{FAlven04}
{Fritze v.~Alvensleben} U.,  2004, in Penetrating bars through masks of cosmic dust : 
the Hubble tuning fork strikes a new note, Proceedings of a conference held at 
Pilanesburg National Park (South Africa). Edited by D. L. Block, I. Puerari, 
K. C. Freeman, R. Groess, and E. K. Block. Astrophysics and space science 
library (ASSL) vol. 319. Dordrecht: Kluwer Academic Publishers, 2004, p.81

\bibitem[\protect\citeauthoryear{{Graham} \& {Guzm{\'a}n}}{{Graham} \&
  {Guzm{\'a}n}}{2003}]{GrahamGuzman03}
{Graham} A.~W.,  {Guzm{\'a}n} R.,  2003, \aj, 125, 2936

\bibitem[\protect\citeauthoryear{{Hao}, {Mao}, {Deng}, {Xia} \& {Wu}}{{Hao}
  et~al.}{2006}]{Hao06+}
{Hao} C.~N.,  {Mao} S.,  {Deng} Z.~G.,  {Xia} X.~Y.,    {Wu} H.,  2006, 
   \mnras, 370, 1339

\bibitem[\protect\citeauthoryear{{Hubble}}{{Hubble}}{1936}]{Hubble36}
{Hubble} E.~P.,  1936, Realm of the Nebulae, Yale University Press

\bibitem[\protect\citeauthoryear{{Illingworth}}{{Illingworth}}{1977}]{Illingwo%
rth77}
{Illingworth} G.,  1977, \apjl, 218, L43

\bibitem[\protect\citeauthoryear{{Jaffe}}{{Jaffe}}{1983}]{Jaffe83}
{Jaffe} W.,  1983, \mnras, 202, 995

\bibitem[\protect\citeauthoryear{{Jedrzejewski}}{{Jedrzejewski}}{1987}]{Jedr87}
{Jedrzejewski} R.~I.,  1987, \mnras, 226, 747

\bibitem[\protect\citeauthoryear{{Jorgensen} \& {Franx}}{{Jorgensen} \&
  {Franx}}{1994}]{JorgensenFranx94}
{Jorgensen} I.,  {Franx} M.,  1994, \apj, 433, 553

\bibitem[\protect\citeauthoryear{{Kissler-Patig} \& {Gebhardt}}{{Kissler-Patig}
  \& {Gebhardt}}{1998}]{Kissler98+}
{Kissler-Patig} M.,  {Gebhardt} K.,  1998, \aj, 116, 2237

\bibitem[\protect\citeauthoryear{{Kormendy} \& {Bender}}{{Kormendy} \&
  {Bender}}{1996}]{KB96}
{Kormendy} J.,  {Bender} R.,  1996, \apjl, 464, L119+

\bibitem[\protect\citeauthoryear{{Krajnovi{\'c}}, {Cappellari}, {de Zeeuw} \&
  {Copin}}{{Krajnovi{\'c}} et~al.}{2006}]{Krajnovic+06}
{Krajnovi{\'c}} D.,  {Cappellari} M.,  {de Zeeuw} P.~T.,    {Copin} Y.,  2006,
  \mnras, 366, 787

\bibitem[\protect\citeauthoryear{{Kuntschner}, {Emsellem}, {Bacon}, {Bureau},
  {Cappellari}, {Davies}, {de Zeeuw}, {Falc{\'o}n-Barroso}, {Krajnovi{\'c}},
  {McDermid}, {Peletier} \& {Sarzi}}{{Kuntschner} et~al.}{2006}]{PaperVI}
{Kuntschner} H.,  {Emsellem} E.,  {Bacon} R.,  {Bureau} M.,  {Cappellari} M.,
  {Davies} R.~L.,  {de Zeeuw} P.~T.,  {Falc{\'o}n-Barroso} J.,  {Krajnovi{\'c}}
  D.,  {McDermid} R.~M.,  {Peletier} R.~F.,    {Sarzi} M.,  2006, \mnras, 369,
  497 (Paper~VI)

\bibitem[\protect\citeauthoryear{{Lauer}}{{Lauer}}{1985}]{Lauer85}
{Lauer} T.~R.,  1985, \mnras, 216, 429

\bibitem[\protect\citeauthoryear{{Lauer}, {Faber}, {Gebhardt}, {Richstone},
  {Tremaine}, {Ajhar}, {Aller}, {Bender}, {Dressler}, {Filippenko}, {Green},
  {Grillmair}, {Ho}, {Kormendy}, {Magorrian}, {Pinkney} \& {Siopis}}{{Lauer}
  et~al.}{2005}]{Lauer+05}
{Lauer} T.~R.,  {Faber} S.~M.,  {Gebhardt} K.,  {Richstone} D.,  {Tremaine} S.,
   {Ajhar} E.~A.,  {Aller} M.~C.,  {Bender} R.,  {Dressler} A.,  {Filippenko}
  A.~V.,  {Green} R.,  {Grillmair} C.~J.,  {Ho} L.~C.,  {Kormendy} J.,
  {Magorrian} J.,  {Pinkney} J.,    {Siopis} C.,  2005, \aj, 129, 2138

\bibitem[\protect\citeauthoryear{{McDermid}, {Emsellem}, {Shapiro}, {Bacon},
  {Bureau}, {Cappellari}, {Davies}, {de Zeeuw}, {Falc{\'o}n-Barroso},
  {Krajnovi{\'c}}, {Kuntschner}, {Peletier} \& {Sarzi}}{{McDermid}
  et~al.}{2006}]{PaperVIII}
{McDermid} R.~M.,  {Emsellem} E.,  {Shapiro} K.~L.,  {Bacon} R.,  {Bureau} M.,
  {Cappellari} M.,  {Davies} R.~L.,  {de Zeeuw} T.,  {Falc{\'o}n-Barroso} J.,
  {Krajnovi{\'c}} D.,  {Kuntschner} H.,  {Peletier} R.~F.,    {Sarzi} M.,
  2006, \mnras, 373, 906 (Paper~VIII)

\bibitem[\protect\citeauthoryear{{Mei}, {Blakeslee}, {Tonry}, {Jord{\'a}n},
  {Peng}, {C{\^o}t{\'e}}, {Ferrarese}, {West}, {Merritt} \&
  {Milosavljevi{\'c}}}{{Mei} et~al.}{2005}]{Mei+05}
{Mei} S.,  {Blakeslee} J.~P.,  {Tonry} J.~L.,  {Jord{\'a}n} A.,  {Peng} E.~W.,
  {C{\^o}t{\'e}} P.,  {Ferrarese} L.,  {West} M.~J.,  {Merritt} D.,
  {Milosavljevi{\'c}} M.,  2005, \apj, 625, 121

\bibitem[\protect\citeauthoryear{{Merritt}}{{Merritt}}{2006}]{Merritt06}
{Merritt} D.,  2006, \apj, 648, 976

\bibitem[\protect\citeauthoryear{{Michard}}{{Michard}}{1994}]{Michard94}
{Michard} R.,  1994, \aap, 288, 401

\bibitem[\protect\citeauthoryear{{Naab}, {Jesseit} \& {Burkert}}{{Naab}
  et~al.}{2006}]{NJB06}
{Naab} T.,  {Jesseit} R.,    {Burkert} A.,  2006, \mnras, 372, 839

\bibitem[\protect\citeauthoryear{{Naab}, {Khochfar} \& {Burkert}}{{Naab}
  et~al.}{2006}]{Naab06+}
{Naab} T.,  {Khochfar} S.,    {Burkert} A.,  2006, \apjl, 636, L81

\bibitem[\protect\citeauthoryear{{Paturel}, {Petit}, {Prugniel}, {Theureau},
  {Rousseau}, {Brouty}, {Dubois} \& {Cambr{\'e}sy}}{{Paturel}
  et~al.}{2003}]{Patu03}
{Paturel} G.,  {Petit} C.,  {Prugniel} P.,  {Theureau} G.,  {Rousseau} J.,
  {Brouty} M.,  {Dubois} P.,    {Cambr{\'e}sy} L.,  2003, \aap, 412, 45

\bibitem[\protect\citeauthoryear{{Peebles}}{{Peebles}}{1969}]{Peebles69}
{Peebles} P.~J.~E.,  1969, \apj, 155, 393

\bibitem[\protect\citeauthoryear{{Ravindranath}, {Ho}, {Peng}, {Filippenko} \&
  {Sargent}}{{Ravindranath} et~al.}{2001}]{Ravindranath+01}
{Ravindranath} S.,  {Ho} L.~C.,  {Peng} C.~Y.,  {Filippenko} A.~V.,
  {Sargent} W.~L.~W.,  2001, \aj, 122, 653

\bibitem[\protect\citeauthoryear{{Rest}, {van den Bosch}, {Jaffe}, {Tran},
  {Tsvetanov}, {Ford}, {Davies} \& {Schafer}}{{Rest} et~al.}{2001}]{Rest+01}
{Rest} A.,  {van den Bosch} F.~C.,  {Jaffe} W.,  {Tran} H.,  {Tsvetanov} Z.,
  {Ford} H.~C.,  {Davies} J.,    {Schafer} J.,  2001, \aj, 121, 2431

\bibitem[\protect\citeauthoryear{{Rix}, {Franx}, {Fisher} \&
  {Illingworth}}{{Rix} et~al.}{1992}]{Rix+92}
{Rix} H.-W.,  {Franx} M.,  {Fisher} D.,    {Illingworth} G.,  1992, \apjl, 400,
  L5

\bibitem[\protect\citeauthoryear{{Rix} \& {White}}{{Rix} \&
  {White}}{1990}]{RixWhite90}
{Rix} H.-W.,  {White} S.~D.~M.,  1990, \apj, 362, 52

\bibitem[\protect\citeauthoryear{{Robertson}, {Bullock}, {Cox}, {Di Matteo},
  {Hernquist}, {Springel} \& {Yoshida}}{{Robertson}
  et~al.}{2006}]{Robertson+06b}
{Robertson} B.,  {Bullock} J.~S.,  {Cox} T.~J.,  {Di Matteo} T.,  {Hernquist}
  L.,  {Springel} V.,    {Yoshida} N.,  2006, \apj, 645, 986

\bibitem[\protect\citeauthoryear{{Robertson}, {Cox}, {Hernquist}, {Franx},
  {Hopkins}, {Martini} \& {Springel}}{{Robertson} et~al.}{2006}]{Robertson+06a}
{Robertson} B.,  {Cox} T.~J.,  {Hernquist} L.,  {Franx} M.,  {Hopkins} P.~F.,
  {Martini} P.,    {Springel} V.,  2006, \apj, 641, 21

\bibitem[\protect\citeauthoryear{{Rubin}, {Graham} \& {Kenney}}{{Rubin}
  et~al.}{1992}]{Rubin+92}
{Rubin} V.~C.,  {Graham} J.~A.,    {Kenney} J.~D.~P.,  1992, \apjl, 394, L9

\bibitem[\protect\citeauthoryear{{Ryden}, {Terndrup}, {Pogge} \&
  {Lauer}}{{Ryden} et~al.}{1999}]{Ryden+99}
{Ryden} B.~S.,  {Terndrup} D.~M.,  {Pogge} R.~W.,    {Lauer} T.~R.,  1999,
  \apj, 517, 650

\bibitem[\protect\citeauthoryear{{Sandage}}{{Sandage}}{2004}]{Sandage04}
{Sandage} A.,  2004, in Penetrating bars through masks of cosmic dust : 
the Hubble tuning fork strikes a new note, Proceedings of a conference held 
at Pilanesburg National Park (South Africa). Edited by D. L. Block, I. Puerari, 
 K. C. Freeman, R. Groess, and E. K. Block. Astrophysics and space science library 
 (ASSL) vol. 319. Dordrecht: Kluwer Academic Publishers, 2004, p.39

\bibitem[\protect\citeauthoryear{{Sandage} \& {Bedke}}{{Sandage} \&
  {Bedke}}{1994}]{Carnegie94}
{Sandage} A.,  {Bedke} J.,  1994, {The Carnegie atlas of galaxies}.
Washington, DC: Carnegie Institution of Washington with The Flintridge
  Foundation, |c1994

\bibitem[\protect\citeauthoryear{{Sarzi}, {Falc{\'o}n-Barroso}, {Davies},
  {Bacon}, {Bureau}, {Cappellari}, {de Zeeuw}, {Emsellem}, {Fathi},
  {Krajnovi{\'c}}, {Kuntschner}, {McDermid} \& {Peletier}}{{Sarzi}
  et~al.}{2006}]{PaperV}
{Sarzi} M.,  {Falc{\'o}n-Barroso} J.,  {Davies} R.~L.,  {Bacon} R.,  {Bureau}
  M.,  {Cappellari} M.,  {de Zeeuw} P.~T.,  {Emsellem} E.,  {Fathi} K.,
  {Krajnovi{\'c}} D.,  {Kuntschner} H.,  {McDermid} R.~M.,    {Peletier} R.~F.,
   2006, \mnras, 366, 1151 (Paper~V)

\bibitem[\protect\citeauthoryear{{Sembach} \& {Tonry}}{{Sembach} \&
  {Tonry}}{1996}]{Sembach96+}
{Sembach} K.~R.,  {Tonry} J.~L.,  1996, \aj, 112, 797

\bibitem[\protect\citeauthoryear{{Somerville} \& {Primack}}{{Somerville} \&
  {Primack}}{1999}]{Somerville+99}
{Somerville} R.~S.,  {Primack} J.~R.,  1999, \mnras, 310, 1087

\bibitem[\protect\citeauthoryear{{Spitzer} \& {Baade}}{{Spitzer} \&
  {Baade}}{1951}]{SB51}
{Spitzer} L.~J.,  {Baade} W.,  1951, \apj, 113, 413

\bibitem[\protect\citeauthoryear{{Statler}}{{Statler}}{2001}]{Statler01}
{Statler} T.~S.,  2001, \aj, 121, 244

\bibitem[\protect\citeauthoryear{{Tonry}, {Dressler}, {Blakeslee}, {Ajhar},
  {Fletcher}, {Luppino}, {Metzger} \& {Moore}}{{Tonry} et~al.}{2001}]{Tonry+01}
{Tonry} J.~L.,  {Dressler} A.,  {Blakeslee} J.~P.,  {Ajhar} E.~A.,  {Fletcher}
  A.~B.,  {Luppino} G.~A.,  {Metzger} M.~R.,    {Moore} C.~B.,  2001, \apj,
  546, 681

\bibitem[\protect\citeauthoryear{{Trujillo}, {Erwin}, {Asensio Ramos} \&
  {Graham}}{{Trujillo} et~al.}{2004}]{Trujillo04+}
{Trujillo} I.,  {Erwin} P.,  {Asensio Ramos} A.,    {Graham} A.~W.,  2004, \aj,
  127, 1917

\bibitem[\protect\citeauthoryear{{Tully}}{{Tully}}{1988}]{Tully88}
{Tully} R.~B.,  1988, Nearby galaxies catalog.
Cambridge and New York, Cambridge University Press, 1988, 221 p.

\bibitem[\protect\citeauthoryear{{Turnbull}, {Bridges} \& {Carter}}{{Turnbull}
  et~al.}{1999}]{Turnbull+99}
{Turnbull} A.~J.,  {Bridges} T.~J.,    {Carter} D.,  1999, \mnras, 307, 967

\bibitem[\protect\citeauthoryear{{van den Bergh}}{{van den
  Bergh}}{1990}]{vdB90}
{van den Bergh} S.,  1990, \apj, 348, 57

\bibitem[\protect\citeauthoryear{{van den Bosch}, {Abel}, {Croft}, {Hernquist}
  \& {White}}{{van den Bosch} et~al.}{2002}]{vdB02+}
{van den Bosch} F.~C.,  {Abel} T.,  {Croft} R.~A.~C.,  {Hernquist} L.,
  {White} S.~D.~M.,  2002, \apj, 576, 21

\bibitem[\protect\citeauthoryear{{van der Marel} \& {Franx}}{{van der Marel} \&
  {Franx}}{1993}]{vdMF93}
{van der Marel} R.~P.,  {Franx} M.,  1993, \apj, 407, 525

\bibitem[\protect\citeauthoryear{{van Driel}, {Arnaboldi}, {Combes} \&
  {Sparke}}{{van Driel} et~al.}{2000}]{vanDriel00+}
{van Driel} W.,  {Arnaboldi} M.,  {Combes} F.,    {Sparke} L.~S.,  2000, \aaps,
  141, 385

\bibitem[\protect\citeauthoryear{{Vitvitska}, {Klypin}, {Kravtsov}, {Wechsler},
  {Primack} \& {Bullock}}{{Vitvitska} et~al.}{2002}]{Vivitska02+}
{Vitvitska} M.,  {Klypin} A.~A.,  {Kravtsov} A.~V.,  {Wechsler} R.~H.,
  {Primack} J.~R.,    {Bullock} J.~S.,  2002, \apj, 581, 799

\bibitem[\protect\citeauthoryear{{Wernli}, {Emsellem} \& {Copin}}{{Wernli}
  et~al.}{2002}]{WEC02}
{Wernli} F.,  {Emsellem} E.,    {Copin} Y.,  2002, \aap, 396, 73

\bibitem[\protect\citeauthoryear{{White}}{{White}}{1984}]{White84}
{White} S.~D.~M.,  1984, \apj, 286, 38

\bibitem[\protect\citeauthoryear{{Young}}{{Young}}{2005}]{Young+05}
{Young} L.~M.,  2005, \apj, 634, 258

\end{thebibliography}

\appendix
%
%
\section{The angular momentum of galaxies and $\lambda_R$}
\label{App:lambda}

The criterion we wish to define should clearly separate the slow and
fast rotators. The first guess is to take something equivalent to the total
angular momentum, which should be expressed as something like $\langle X |V|\rangle$
where $X$ is the distance to the spin axis. To avoid having to determine that
axis, we chose $\langle R |V| \rangle$, where $R$ is the distance to the
centre. The absolute value of $V$ is used as we are interested in the presence
of local streaming motion (e.g. see NGC~4550).
This is then naturally normalised by $\langle X \sqrt{V^2 + \sigma^2} \rangle$ where $V^2 + \sigma^2$ is the second
order velocity moment.

Another route would be to get closer to some physical quantity such as
the angular momentum per unit mass. This can be represented via the spin
parameter \cite[e.g.][]{Peebles69}:
\begin{equation}
\lambda = \frac{J \sqrt{\left|E\right|}}{G M^{2.5}}\, ,
\end{equation}
where $J$, $E$, and $M$ are the total angular momentum, energy and mass, and $G$
is Newton's constant of gravity. The derivation of $\lambda$ in principle requires an accurate
distance for each galaxy. Since we wish to stay close to our \Sauron\
measurements we need to define something which can be readily
evaluated. This is possible by first rewriting $\lambda$ as:
\begin{equation}
\lambda = \frac{J/M \sqrt{\left|E\right|/M}}{G M}\, .
\end{equation}
For two-dimensional integral-field data, we can approximate the spin
parameter by using the scalar virial relations and sky averaging over the
field-of-view, weighted with the surface brightness. The scalar virial relations
are $2E=-2K=W$, with the total kinetic and potential
energy related to the mean-square speed of the system's stars $\VV$
respectively as $2K/M=\VV^2$ and $W/M=-GM/r_g$. Here, $r_g$ is the
gravitational radius, which can be related to the half-mass radius
$r_h$ given the mass model of the stellar system.  For e.g. a
spherical symmetric \cite{Jaffe83} model, we have $r_g=2\,r_h$. 

When relating $r_g$ (or $r_h$) to the observed effective radius $R_e$,
we have to take into account the projection as a function of the viewing
angle of the stellar system which in general is not spherical.
Furthermore, while $r_g$ and $r_h$ are related to the mass
distribution, $R_e$ depends on the light distribution. Therefore,
given $r_g=\kappa_R\,R_e$ the conversion factor $\kappa_R$ is a
function of the mass model, viewing direction and mass-to-light ratio.
In a similar way a conversion factor is needed in the approximation of
the total angular momentum by the observed radius times velocity:
$J/M=\kappa_J\,R|V|$ \citep[see][]{Franx88}. Furthermore, 
while $\VV^2$ is the \textit{total} mean-square speed, 
we only observe velocity moments projected along the line-of-sight, 
e.g. $V$ and $\sigma$. In order to retrieve $\VV^2$, 
we should therefore allow multiplying factors
$\kappa_V$ and $\kappa_S$ associated with the corresponding observed $V^2$ and $\sigma^2$
(in the case of an isothermal sphere, $\kappa_V = 1$ and  $\kappa_S=3$).

We thus find the following expressions:
\begin{equation}
  J/M = \kappa_J \, \langle R|V| \rangle\, ,
\end{equation}
\begin{equation}
  2E/M = - \, \langle \kappa_V \, V^2 + \kappa_S \, \sigma^2 \rangle\, ,
\end{equation}
\begin{equation}
  GM  = \kappa_R\, \langle R(\kappa_V \, V^2+ \kappa_S \, \sigma^2) \rangle\, ,
\end{equation}
resulting in an approximate spin parameter
\begin{equation}
  \label{eq:lamapprox}
  \lambda \sim \frac{\kappa_J}{\kappa_R\sqrt{2}} \, 
  \frac{\langle R |V| \rangle \, \sqrt{\langle \kappa_V \, V^2 + \kappa_S \, \sigma^2 \rangle}}
{\langle R \left(\kappa_V \, V^2 + \kappa_S \, \sigma^2\right) \rangle}\, .
\end{equation}
For simplicity, we will use the
observed second order velocity moment $V^2 + \sigma^2$, so 
that both $\kappa_V = 1$ and $\kappa_S=1$, and thus define $\lambda_f$ as:
\begin{equation}
\lambda_f \equiv \frac{\langle R |V| \rangle \, \sqrt{\langle V^2 + \sigma^2 \rangle}}
{\langle R \left(V^2 + \sigma^2\right) \rangle}\, .
\end{equation}
We finally approximate $\lambda_f$ with $\lambda_R$ by writing
\begin{equation}
\lambda_f \sim \lambda_R \equiv \frac{\langle R |V| \rangle}{\langle R  \sqrt{ V^2 + \sigma^2 }\rangle}\, .
\end{equation}
When deriving both the values of $\lambda_f$ and $\lambda_R$ for the 48 E and S0 \Sauron\ galaxies, we 
indeed find a tight relation:
\begin{equation}
\lambda_R = (0.95\pm0.04) \times \lambda_f\, .
\end{equation}
This relation is in fact also valid for the two-integral models as derived 
in Appendix~\ref{App:2I}, so that we expect $\lambda_R$ to be roughly equal to
(though slightly smaller than) $\lambda_f$.
For typical values of the conversion factors, $\kappa_J=2$,
$\kappa_R=3$, we therefore find that $\lambda \sim \sqrt{2}/3 \, \lambda_R$.

An alternative to $\lambda_R$ is the unbounded ${\lambda'_R} =  \langle R |V| \rangle 
/ \langle R  \sigma\rangle$. In the
context of early-type galaxies we favour $\lambda_R$ which includes a
mass-like normalisation by the second order velocity moment $V^2 + \sigma^2$. For
isotropic (two-integral) models, this implies $\lambda_R \sim
\sqrt{\epsilon}$.
For slow rotators $\lambda_R \sim {\lambda'}_R$, since these galaxies
have by definition small mean stellar velocities.

\section{Two-integral models}
\label{App:2I}

In order to examine the behaviour of $\lambda_R$ in more detail, we constructed two-integral dynamical 
Jeans models for 7 \Sauron\ galaxies using the MGE formalism, namely NGC\,524, 3377, 4459,
4486, 4552, 4621, and 5813. Details on the assumptions and resulting mass models 
for these galaxies can be found in Paper~IV. For each galaxy, we assumed 8 different inclinations
(5, 15, 25, 35, 50, 65, 80, and 90\degr, the latter corresponding to an edge-on view),
and derived the first two line-of-sight velocity moments up to 1~$R_e$.

These models were used to obtain measurements of $\lambda_R$, $\lambda_f$ and $V/\sigma$,
to determine whether these three quantities behave similarly with respect to basic
parameters such as the inclination or anisotropy parameter.
We first obtain that $\lambda_R = (0.93 \pm 0.04) \times \lambda_f$, 
a correlation very similar to what
was found for the observed 48 \Sauron\ galaxies (see Appendix~\ref{App:lambda}).
The variation of $\lambda_R$ with inclination follows that of $V/\sigma$ 
within better than 10\% for two-integral models. These results do not depend on the 
global anisotropy, as long as the internal dynamics of the model is fixed 
before changing the viewing angle. A radial variation in the anisotropy 
profile can, however, significantly change this situation, as
$\lambda_R$ includes an additional explicit dependence on radius.
Assuming isotropy for all 7 MGE Jeans models, we observe a tight correlation
between $\lambda_R$ and $V/\sigma$.
A simple approximation for the relation between $\lambda_R$ 
and $V/\sigma$ can then be obtained by introducing a scaling factor $\kappa$:
\begin{equation}
\lambda_R = \frac{\langle R V \rangle}{\langle R  \sqrt{ V^2 + \sigma^2 }\rangle}
\approx \frac{\kappa \, \left( V / \sigma \right)}{\sqrt{1 + \kappa^2 \, \left( V / \sigma \right)^2}}
\, .
\label{eq:kappa}
\end{equation}
Using our two-integral models we find a best fit relation with 
$\kappa = 1.2 \pm 0.1$, slightly
higher than but still consistent with the best fit 
value $\kappa = 1.1 \pm 0.1$ obtained using our 48 \Sauron\ Es and S0s.
We can in fact directly solve for $\kappa$ in Eq.~(\ref{eq:kappa}) in terms
of the values of $\lambda_R$ and $V/\sigma$ measured for individual
two-integral models: there is in fact a rather large spread in the values of
$\kappa$, which ranges from 1 and 1.4, depending on the galaxy. 
This shows that non-homology is an important 
driver of changes in $\lambda_R$ at constant $V / \sigma$.

\begin{figure}
\centering
\epsfig{file=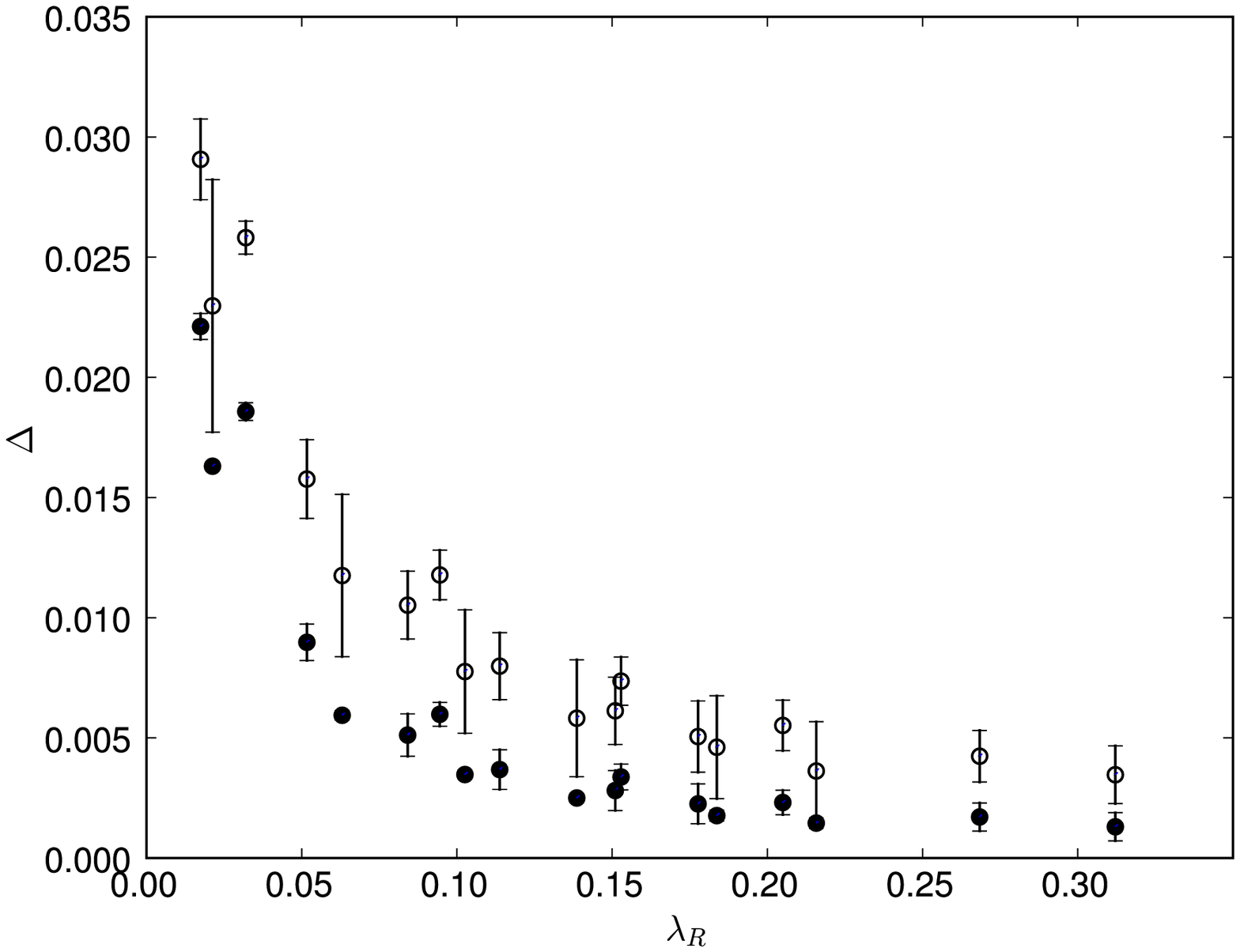, width=\columnwidth}
\caption{Effect of noise in the kinematics on the estimate of $\lambda_R$ (filled symbols)
and $V/\sigma$ (empty symbols). $\Delta$, the difference between the measured value (with noise) 
and the expected one (noiseless), is plotted against $\lambda_R$.
The input noise has been set to 10~\kms\ for both $V$ and $\sigma$.}
\label{fig:erreffect}
\end{figure}

Another important issue when deriving $\lambda_R$ and $V/\sigma$ comes from the presence
of noise in the measurements. When $V$ is sufficiently large, the random errors in the
kinematic measurements will affect both $\lambda_R$ and $V/\sigma$ randomly.
As $V$ becomes small in absolute value, $\langle V^2\rangle$ and
$\langle R \, |V|\rangle$ become increasingly biased (being artificially increased), which affects 
estimates of $\lambda_R$ and $V/\sigma$. In Fig.~\ref{fig:erreffect}, we show the 
effect of a random error of 10~\kms\ in both $V$ and $\sigma$ (typical in the outer
part of \Sauron\ kinematic maps; see Paper~III) 
on both $\lambda_R$ and $V/\sigma$, in terms of $\lambda_R$.
$\Delta$ is the difference between the measured value (with noise) 
and the expected one (noiseless) and is plotted with respect to the true $\lambda_R$ value.
The effect is in general relatively small, of the order of 7\% on $\lambda_R$ and
almost 15\% on $V/\sigma$, for $\lambda_R \sim 0.1$. 
When $V$ is close to zero everywhere, as is the case for the three \Sauron\ galaxies with the lowest
$\lambda_R$ values, the presence of noise yields an overestimate of
about 0.02 in $\lambda_R$ (depending on the velocity dispersion values), an effect consistent
with the observed values for these galaxies (see Fig.~\ref{fig:RVS}).
As expected, Fig.~\ref{fig:erreffect} also shows that $\lambda_R$ is in general
about a factor of two less sensitive to errors in the kinematic measurements than $V/\sigma$.

\label{lastpage}

\end{document}